\documentclass[aip,cha,
reprint,
secnumarabic,
nofootinbib, tightenlines,
nobibnotes, showkeys, showpacs,
superscriptaddress,
]{revtex4-1}

                        \newif\ifpaper \newif\ifPDF               %%
                        \newif\ifOUP \newif\ifdraft               %%
                        \newif\ifdasbuch                          %%
                        \newif\ifsolutions \newif\ifblog          %%
                        \blogtrue                                 %%
                        \draftfalse       %% commented, WWW/drafts %%
                        \dasbuchtrue %% DasBuch, not QFT lectures %%
                        \solutionstrue %% include solutions       %%
                        \paperfalse\PDFtrue %% hyperlinked        %%
                        \OUPfalse                %% ChaosBook %%%%%%

\usepackage{float}
\usepackage{natbib}
\usepackage{amsmath,amsfonts,amssymb,amsbsy,amscd,amsgen}
\usepackage{dcolumn}% Align table columns on decimal point
\usepackage{bm}% bold math
\usepackage[pdftex]{graphicx}
\usepackage{url}
\usepackage{subfigure}
\usepackage{ifthen}

\usepackage{fancyhdr}
\graphicspath{{figs/}}  %% directories with graphics files
\usepackage[pdftex,colorlinks]{hyperref} %% hyperlinks, LAST package called

\ifpaper % prepare for B&W paper printing:
       
       \renewcommand{\color}[1]{}       % B&W
       \newcommand{\wwwcb}[1]{{\tt ChaosBook.org#1}}
       \newcommand{\arXiv}[1]{ {\tt arXiv:#1}}
\else % prepare hyperlinked pdf
        \newcommand{\wwwcb}[1]{       % keep homepage flexible:
                  {\tt \href{http://ChaosBook.org#1}
              {ChaosBook.org#1}}}

       \newcommand{\arXiv}[1]{
              {\tt \href{http://arXiv.org/abs/#1}{arXiv:#1}}}
\fi

\ifpaper % prepare for B&W paper printing:
\else % prepare hyperlinked pdf
\fi %%%% prepare for B&W paper printing END %%%%%%

\newcommand{\Mvar}{\ensuremath{{A}}}  % stability matrix
\newcommand{\jMps}{\ensuremath{{J}}}   % jacobian matrix, phase space/state space
\newcommand{\monodromy}{\ensuremath{{M}}}   % monodromy matrix, full Poincare cut
\newcommand{\Lg}{\ensuremath{{T}}}   % Predrag Lie algebra generator
\newcommand{\matrixRep}{\ensuremath{{D}}}  %  matrix rep of a group element
\newcommand{\LieEl}{\ensuremath{g}}  %  a group element, often replaced by \matrixRep
\newcommand{\fFslice}{first Fourier mode slice}
\newcommand{\FFslice}{First Fourier mode slice}

\newcommand{\beq}{\begin{equation}}
\newcommand{\continue}{\nonumber \\ }
\newcommand{\nnu}{\nonumber}
\newcommand{\eeq}{\end{equation}}
\newcommand{\ee}[1] {\label{#1} \end{equation}}
\newcommand{\bea}{\begin{eqnarray}}
\newcommand{\ceq}{\nonumber \\ & & }
\newcommand{\eea}{\end{eqnarray}}

\newcommand{\rf}     [1] {~\cite{#1}}
\newcommand{\refref} [1] {ref.~\cite{#1}}

\newcommand{\refrefs}[1] {refs.~\cite{#1}}

\newcommand{\refeq}  [1] {(\ref{#1})}

\newcommand{\reffig} [1] {figure~\ref{#1}}

\newcommand{\refFig} [1] {Figure~\ref{#1}}

\newcommand{\reftab} [1] {Table~\ref{#1}}
\newcommand{\refTab} [1] {Table~\ref{#1}}
\newcommand{\reftabs}[2] {Tables~\ref{#1} and~\ref{#2}}
\newcommand{\refsect}[1] {Section~\ref{#1}}

\newcommand{\refappe}[1] {appendix~\ref{#1}}

\newcommand{\refAppe}[1] {Appendix~\ref{#1}}

\renewcommand{\reffig} [1] {Fig.~\ref{#1}}

\renewcommand{\refFig} [1] {Fig.~\ref{#1}}

\renewcommand{\refref} [1] {Ref.~\onlinecite{#1}}

\renewcommand{\refrefs}[1] {Refs.~\onlinecite{#1}}

\newcommand{\etc}{{etc.}}       % APS
\newcommand{\etal}{{\em et al.}}    % etal in italics, APS too
\newcommand{\ie}{{i.e.}}        % APS
\newcommand{\po}{periodic orbit}
\newcommand{\Po}{Periodic orbit}
\newcommand{\rpo}{rela\-ti\-ve periodic orbit}
\newcommand{\Rpo}{Rela\-ti\-ve periodic orbit}
\newcommand{\eqv}{equi\-lib\-rium}

\newcommand{\eqva}{equi\-lib\-ria}
\newcommand{\Eqva}{Equi\-lib\-ria}
\newcommand{\reqv}{rela\-ti\-ve equi\-lib\-rium}

\newcommand{\reqva}{rela\-ti\-ve equi\-lib\-ria}

\newcommand{\reducedsp}{reduced state space}

\newcommand{\cycForm}{cycle averaging formula}

\newcommand{\evOper}{evolution oper\-ator}

\newcommand{\FPoper}{Perron-Frobenius oper\-ator} % Pesin's ordering

\newcommand{\statesp}{state space}

\newcommand{\stabmat}{stability matrix}     % stability matrix, velocity gradients
\newcommand{\Fd}{spec\-tral det\-er\-min\-ant}
\newcommand{\dmn}{-dimensional}  %  experimental 220ct2009
\newcommand{\KS}{Kuramoto-Siva\-shin\-sky}

\newcommand{\cLf}{complex Lorenz flow}

\newcommand{\slice}{slice}

\newcommand{\mslices}{method of slices}
\newcommand{\Mslices}{Method of slices}

\newcommand{\template}{template} % {slice-fixing point} % {reference state}
\newcommand{\sliceBord}{slice border}

\newcommand{\Un}[1]{\ensuremath{\textrm{U}(#1)}}         % in DasBuch
\newcommand{\On}[1]{\ensuremath{\textrm{O}(#1)}}
\newcommand{\SOn}[1]{\ensuremath{\textrm{SO}(#1)}}         % in DasBuch

\renewcommand\Im{{\rm Im\,}}
\renewcommand\Re{{\rm Re\,}}
\newcommand{\braket}[2]
		   {\langle{#1}\vphantom{#2}|\vphantom{#1}{#2}\rangle}
\newcommand{\transp}[1]{{#1}{}^\top}
\newcommand{\norm}[1]{\left\Arrowvert \, #1 \, \right\Arrowvert}
\renewcommand{\det}{\mbox{\rm det}\,}
\newcommand{\tr}{\mbox{\rm tr}\,}
\newcommand{\pS}{\ensuremath{{\cal M}}}          % symbol for state space
\newcommand{\ssp}{\ensuremath{x}}                % state space point

\newcommand{\EQV}[1]{\ensuremath{EQ_{#1}}} %experimental
\newcommand{\REQV}[2]{\ensuremath{TW_{#1#2}}} % #1 is + or -
\newcommand\stagn{q}      %equilibrium/stagnation point suffix
\newcommand{\rpprime}{{\tilde{p}}}  % relative periodic prime orbit

\newcommand{\Lop}{\ensuremath{{\cal L}}}       % evolution operator
\newcommand{\Aop}{\ensuremath{{\cal A}}}       % evolution generator
\newcommand{\matId}{\ensuremath{{\bf 1}}}      % matrix identity

\newcommand{\obser}{\ensuremath{\omega}}     % an observable from phase space to R^n
\newcommand{\Obser}{\ensuremath{\Omega}}     % time integral of an observable
\newcommand{\timeAver} [1]{\overline{#1}}
\newcommand{\pde}{\partial}
\newcommand{\reals}{\mathbb{R}}

\newcommand\period[1]{{\ensuremath{T_{#1}}}}         %continuous cycle period
\newcommand{\cl}[1]{{\ensuremath{n_{#1}}}}   % discrete length of a cycle, Predrag

\newcommand\flow[2]{{f^{#1}(#2)}}
\newcommand{\vel}{\ensuremath{{v}}}   % state space velocity
\newcommand{\MvarRed}{\ensuremath{\hat{\Mvar}}}  % stability matrix
\newcommand{\jMpsRed}{\ensuremath{\hat{\jMps}}}   % jacobian matrix, symmetry reduced space/state space
\newcommand{\ExpaEig}{\ensuremath{\Lambda}}
\newcommand{\Lyap}{\ensuremath{\lambda}}            %Lyapunov exponent
\newcommand{\monodromyRed}{\ensuremath{\hat{\monodromy}}}   % monodromy matrix, full Poincare cut
\newcommand{\oneMinJ}[1]
           {\left|\det\!\left(\matId-\monodromy_p^{#1}\right)\right|}
\newcommand{\oneMinJred}[1]
           {\left|\det\!\left(\matId-\monodromyRed_p^{#1}\right)\right|}
\newcommand{\eigExp}[1][]{
     \ifthenelse{\equal{#1}{}}{\ensuremath{\lambda}}{\ensuremath{\lambda^{(#1)}}}}
\newcommand{\eigRe}[1][]{
     \ifthenelse{\equal{#1}{}}{\ensuremath{\mu}}{\ensuremath{\mu^{(#1)}}}}
\newcommand{\eigIm}[1][]{
     \ifthenelse{\equal{#1}{}}{\ensuremath{\omega}}{\ensuremath{\omega^{(#1)}}}}

\newcommand{\cycle}[1]{{\ensuremath{\overline{#1}}}}
\newcommand{\gSpace}{\ensuremath{{\bf \theta}}}   % MA group rotation parameters
\newcommand{\velRel}{\ensuremath{c}}    % relative state or phase velocity
\newcommand{\phaseVel}{phase velocity}      % pipe slicing
\renewcommand{\ssp}{\ensuremath{a}} % Real valued statespace variable
\renewcommand{\rpprime}{\ensuremath{p}}

\newcommand{\invpol}{\ensuremath{p}} % Invariant polynomials
\newcommand{\cartpt}[1]{\left( #1 \right)} % Points in full Cartesian 4D space
\newcommand{\polpt}[1]{$\left\{ #1 \right\}$} % Points in polar coordinates
\newcommand{\invpt}[1]{\left[ #1 \right]} % Points in invariant polynomial basis

\newcommand{\twoMode}{Two-mode}
\newcommand{\twomode}{two-mode}
\newcommand{\slicePlane}{slice hyperplane}

\newcommand{\pSRed}{\ensuremath{\hat{\cal M}}} % reduced state space Jan 2012
\newcommand{\sspRed}{\ensuremath{\hat{\ssp}}}    % reduced state space point Jan 2012
\newcommand{\velRed}{\ensuremath{\hat{\vel}}}    % ES reduced state space velocity Jan 2012
\newcommand{\slicep}{{\ensuremath{\sspRed'}}}   % slice-fixing point Jan 2012
\newcommand{\sliceTan}[1]{\ensuremath{t'_{#1}}}    % group orbit tangent at slice-fixing
\newcommand{\groupTan}{\ensuremath{t}}    % group orbit tangent
\newcommand{\Group}{\ensuremath{G}}         % Predrag Lie or discrete group

\newcommand{\zeit}{\ensuremath{\tau}}  %time variable
\newcommand{\sspC}{\ensuremath{z}} %Complex valued state space variable
\newcommand{\sspRedC}{\ensuremath{\hat{\sspC}}}
\newcommand{\conf}{\ensuremath{x}} %Configuration space coordinate

\newcommand{\NS}{Navier-Stokes}
\newcommand{\NSe}{Navier-Stokes equations}
\newcommand{\ii}{\ensuremath{\mathrm{i}}} % sqrt{-1}

\begin{document}

\title[Periodic orbit analysis of a system with continuous symmetry]
{Periodic orbit analysis of a system with continuous symmetry - a tutorial}

\author{Nazmi Burak Budanur}
\email{budanur3@gatech.edu}
\affiliation{
 School of Physics and Center for Nonlinear Science,
 Georgia Institute of Technology,
 Atlanta, GA 30332
}
\author{Daniel Borrero-Echeverry}
\affiliation{
 School of Physics and Center for Nonlinear Science,
 Georgia Institute of Technology,
 Atlanta, GA 30332
}
\affiliation{
 Department of Physics,
 Reed College,
 Portland OR 97202
}
\author{Predrag Cvitanovi\'{c}}
\affiliation{
 School of Physics and Center for Nonlinear Science,
 Georgia Institute of Technology,
 Atlanta, GA 30332
}

\date{18 June 2015}

\begin{abstract}
Dynamical systems with translational or rotational symmetry arise
frequently in studies of spatially extended physical systems, such as
Navier-Stokes flows on periodic domains. In these cases, it is natural to
express the state of the fluid in terms of a Fourier series truncated to a
finite number of modes.
Here, we study a 4-dimensional model with chaotic dynamics and
\SOn{2} symmetry similar to those that appear in fluid dynamics problems.
A crucial step in the analysis of such a system
is symmetry reduction. We use the model to illustrate different
symmetry-reduction techniques. Its relative equilibria are conveniently
determined by rewriting the dynamics in terms of a symmetry-invariant
polynomial basis. However, for the analysis of its chaotic dynamics, the `method of
slices', which is applicable to very high-dimensional problems, is
preferable. We show that a Poincar\'e section taken on the `slice' can be
used to further reduce this flow to what is for all practical purposes a
unimodal map. This enables us to systematically determine all relative
periodic orbits and their symbolic dynamics up to any desired period. We
then present cycle averaging formulas adequate for systems with continuous symmetry
and use them to compute dynamical averages using relative periodic orbits. The convergence
of such computations is discussed.
\end{abstract}

\maketitle

\begin{quotation}
Periodic orbit theory provides a way to compute dynamical averages for
chaotic flows by means of {\cycForm s} that relate the time averages of
observables to the spectra of unstable periodic orbits. Standard
{\cycForm s} are valid under the assumption that the stability
multipliers of all periodic orbits have a single marginal direction
corresponding to time evolution and are hyperbolic in all other
directions. However, if a dynamical system has $N$ continuous symmetries,
periodic orbits are replaced by relative periodic orbits, invariant
$(N+1)$-dimensional tori with marginal stability in $(N+1)$ directions.
Such exact invariant solutions arise in studies of turbulent flows, such
as pipe flow or plane Couette flow, which have continuous symmetries.
In practice, the translational invariance of these flows is approximated in numerical
simulations by using periodic domains so that the state of the fluid
is conveniently expressed as a Fourier series, truncated to a large but finite
number (from tens to thousands) of Fourier modes. This paper is a tutorial on
how such problems can be analyzed using periodic orbit theory. We illustrate
all the necessary steps using a simple `\twomode' model as an example.
\end{quotation}

\section{Introduction}
\label{s:intro}

Recent experimental observations of traveling waves in pipe flows have
confirmed the intuition from dynamical systems theory that invariant solutions
of \NSe\ play an important role in shaping the \statesp\ of turbulent
flows.\rf{science04} When one casts fluid flow equations in a
particular basis, the outcome is an infinite dimensional dynamical system
that is often equivariant under transformations such as
translations, reflections and rotations. For example, when periodic
boundary conditions are imposed along the streamwise direction, the equations
for pipe flow retain their form under the action of streamwise translations,
azimuthal rotations and reflections about the central axis, \ie, they are equivariant
under the actions of $\SOn{2}\times\On{2}$. In this case it is natural
to express the state of the fluid in a Fourier basis. However,
as the system evolves, the nonlinear terms in the equations mix the
various modes, so that the state of the system evolves not only along the
symmetry directions, but also along directions transverse to them.
This complicates the dynamics and gives rise to high dimensional coherent
solutions such as \reqva\ and \rpo s, which take on the roles played by
\eqva\ and \po s in flows without symmetry.

There is an extensive literature on equivariant dynamics,
which can be traced back to Poincar\'e's work on the 3-body problem.\rf{Poinc1896}
Early references in the modern dynamical systems
literature that we know of are works of Smale,\rf{Smale70I}
Field,\rf{Field70} and Ruelle.\rf{ruell73} Our goal here is not to
provide a comprehensive review of this literature, or study its
techniques in generality. For those, we refer the reader to monographs
by Golubitsky and Stewart,\rf{golubitsky2002sp} and Field.\rf{Field07}
Our aim here is much more modest: We would like to
provide a hands-on introduction to some of the concepts from
equivariant dynamical systems theory, with an emphasis on those aspects
relevant to the application of the periodic orbit theory to these systems. To this end,
we undertake a step-by-step tutorial approach and illustrate each concept on a
\twomode\ \SOn{2} equivariant normal form that has the minimal
dimensionality required for chaotic dynamics. We provide
visualizations of geometrical concepts, whenever possible.
While the example studied here has no physical significance,
such an analysis should ultimately be applicable to numerical solutions of
turbulent flows on periodic domains, once sufficiently many exact
invariant solutions become numerically accessible.

The rest of the paper is organized as follows: In \refsect{s:symm}, we
define basic concepts and briefly review the relevant symmetry reduction
literature. In \refsect{s:twoMode}, we introduce the \twomode\ model
system, discuss several of its symmetry-reduced representations,
and utilize a symmetry-reduced polynomial representation to find the only \reqv\ of the
system. In \refsect{s:numerics}, we show how the \mslices\ can be used to
quotient the symmetry and reduce the dynamics onto a symmetry-reduced
\statesp\ or `\slice '. A Poincar\'e section taken on the \slice\ then
reduces the 4\dmn\ chaotic dynamics in the full \statesp\ to an approximately
one-dimensional, unimodal Poincar\'e return map. The return map is then
used to construct a finite grammar symbolic dynamics for the flow and
determine {\em all} \rpo s up to a given period. In \refsect{s:DynAvers},
we present {\cycForm s} adequate for systems with continuous symmetries
and use the relative periodic orbits calculated in \refsect{s:numerics}
to calculate dynamically interesting observables. Finally, in \refsect{s:concl},
we discuss possible applications of the \mslices\ to various spatially
extended systems.

The main text is supplemented by two appendices. \refAppe{s:newton} describes
the multi-shooting method used to calculate the \rpo s.
\refAppe{s:schur} discusses how periodic Schur decomposition can be used
to determine their Floquet multipliers, which can differ by 100s
of orders of magnitude even in a model as simple as the \twomode\ system.

\section{Continuous symmetries}
\label{s:symm}

A dynamical system $\dot{\ssp}=\vel(\ssp)$ is said to be
\emph{equivariant} under the group \Group\ of symmetry transformations if
%                                                \toCB
\beq
   \vel( \ssp )
    =  \matrixRep(\LieEl)^{-1}\vel(\matrixRep(\LieEl)\ssp)
   \,
\ee{equiv}
for every point $\ssp$ in the \statesp\ $\pS$ and every element $\LieEl \in
\Group$, where \LieEl\ is an abstract group element and
$\matrixRep(\LieEl)$ is its $[d\!\times\!d]$ matrix representation.
Infinitesimally, the equivariance condition \refeq{equiv} can be expressed as
a vanishing Lie derivative\rf{DasBuch}
%                                                \toCB
\beq
  \Lg \, \vel(\ssp)  - \Mvar(\ssp) \, \groupTan(\ssp) =0
  \,,
\ee{inftmInv}
where
$\Mvar(\ssp)$ is the $[d\!\times\!d]$ \stabmat\, with elements
$\Mvar_{ij}(\ssp)={\pde \vel_i}/{\pde\ssp_j}$, $ \groupTan(\ssp)
= \Lg \ssp $ is the group tangent at $\ssp$, and $\Lg$ is the
$[d\!\times\!d]$ generator of infinitesimal transformations, such that
$\matrixRep(\theta) = \exp(\theta\Lg)$, where the phase $\theta \in [0,2\pi)$
parametrizes the group action. (We shall interchangeably use notations
$\matrixRep(\LieEl)$ and $\matrixRep(\theta)$.) In general, there is a
generator associated with each continuous symmetry. For the simple model
considered here, which has a single $\SOn{2}$ symmetry, there is only one
parameter $\theta$, so we only have one generator \Lg.

If the trajectory of a point $\ssp_\stagn$ coincides with its group
orbit, \ie, for every $\zeit$ there is a group transformation such that
\beq
\ssp (\zeit)
    = \ssp_\stagn + \int_0^\zeit \!\!d\zeit' \vel(\ssp (\zeit'))
    = \matrixRep(\theta (\zeit))\,\ssp_\stagn
  \,,
\ee{releq}
$\ssp_\stagn$ is a point on \emph{\reqv} $\stagn$. In our case, this is a
1-torus in \statesp. Expanding both sides of \refeq{releq} for infinitesimal time
verifies that the group tangent and the velocity vector are parallel, i.e.,
%  need this written out step-by-step in ChaosBook                  \toCB
 $\vel(\ssp_\stagn) = \dot{\theta}(0) \, \groupTan(\ssp_\stagn)$.
By symmetry, this must hold for all $\ssp(\zeit) \in q$, so for \reqva\
the \emph{\phaseVel} is constant, $\dot{\theta}(\zeit) = \velRel$.
Multiplying the equivariance condition \refeq{inftmInv} by $\velRel$, we
find that velocity is a marginal stability eigenvector in the reference frame co-moving
with the \reqv,
\beq
(\Mvar (\ssp) - \velRel \Lg) \vel (\ssp) = 0
\,,\qquad \ssp \in \pS_\stagn
\,.
\ee{ReqvMargEig}

A \statesp\ point $\ssp_\rpprime$ lies on a \emph{\rpo} of period
$\period{\rpprime}$ if its trajectory first intersects its group orbit after
a finite time $\period{\rpprime}$,
%                                                \toCB
\beq
\ssp(\period{\rpprime})
    = \ssp_\rpprime
     + \int_0^\period{\rpprime} \!\!\!\!d\tau' \vel(\ssp (\tau'))
    = \matrixRep(\theta_\rpprime) \,  \ssp_\rpprime
  \,,
\ee{relpo}
with a phase $\theta_{\rpprime}$. In systems with \SOn{2} symmetry,
\rpo s are topologically 2-tori, where the trajectory of
$\ssp_\rpprime$ generically traces out the torus ergodically by
repeating the same path shifted by the group action
$\matrixRep(\theta_\rpprime)$ after each prime period
$\period{\rpprime}$. As we will see in \refsect{s:numerics}, these tori
can be very convoluted and difficult to visualize. In special cases where
$\theta_{\rpprime}=0$, the solution is a \po, a 1-dimensional loop in \statesp\ and
the 2-torus is generated by all actions of the symmetry group on this
loop.

The linear stability of \rpo s is captured by their \emph{Floquet
multipliers}  $\ExpaEig_{p,j}$, the eigenvalues of the Jacobian $\jMpsRed_{\rpprime}$
of the  time-forward map $\ssp(\zeit)=\flow{\zeit}{\ssp(0)}$. $\jMpsRed_{\rpprime}$ is defined as
\beq
\jMpsRed_{\rpprime}
= \matrixRep(- \theta_\rpprime ) \jMps^\period{\rpprime} (\ssp_\rpprime)
\,, \; \mbox{~where~}\;
\jMps^{\zeit}_{ij} (\ssp(0)) = \frac{\partial\ssp_i(\zeit)}{\partial\ssp_j(0)}\, .
\ee{e-rpoJacobian}
The magnitude of $\ExpaEig_{p,j}$ determines whether a small perturbation
along its corresponding eigendirection (or Floquet vector) will expand or
contract after one period. If the magnitude of $\ExpaEig_{p,j}$ is
greater than $1$, the perturbation expands; if it is less than $1$, the
perturbation contracts. In systems with $N$ continuous symmetries, \rpo s
have $(N+1)$ marginal directions ($\left|\ExpaEig_{p,j}\right| = 1$),
which correspond to the temporal evolution of the flow and the $N$
symmetries. By applying symmetry reduction, the marginal Floquet
multipliers corresponding to the symmetries are replaced by $0$, so that
periodic orbit theory, which requires that the flow have only one
marginal direction, becomes applicable.

\emph{Symmetry reduction} is a coordinate transformation that maps
all the points on a group orbit $\matrixRep(\theta) \ssp$, which are
equivalent from a dynamical perspective, to a single representative point
in a symmetry reduced space. After symmetry reduction, \reqva\ and \rpo s
are converted to \eqva\ and \po s in a
reduced \statesp\ without loss of dynamical information; the full \statesp\
trajectory can always be retrieved via a reconstruction equation.

One well-studied technique for symmetry reduction, which
works well for low-dimensional dynamical systems, such as the
Lorenz system, is to recast the dynamical equations in terms of
invariant polynomials\rf{GL-Gil07b}. 
However, there are multiple
difficulties associated with using these techniques. Computing
invariants is a non-trivial problem, and even for the simple case
of \SOn{2}, computer algebra methods for finding invariants become
impractical for systems with more than a dozen
dimensions.\rf{gatermannHab} Moreover, the projection of a linear
equivariant vector field onto orbit space is not necessarily a linear
operation.\rf{Koenig97} This means that even when conducting basic
operations such as the linearization of nonlinear vector fields, special
attention has to be paid to the choice of invariants, even when it is
possible to find them. In contrast, the
\mslices,\rf{rowley_reconstruction_2000,BeTh04,SiCvi10,FrCv11,atlas12,ACHKW11,BudCvi14}
which we study in detail here, is a symmetry reduction scheme applicable
to high-dimensional flows like the \NS\ equations.\rf{WiShCv15}

\subsection{\Mslices}
\label{s-slice}

In a system with $N$ continuous symmetries, a \emph{\slice} \pSRed\
is a codimension $N$ submanifold of \pS\ that cuts every group orbit
once and only once. In the \emph{\mslices}, the solution of a
$d$-\dmn\ dynamical system is represented as a symmetry-reduced
trajectory $\sspRed (\zeit)$ within the $(d-N)$-\dmn\ \slice\ and $N$
time dependent group parameters $\theta(\zeit)$, which map $\sspRed
(\zeit)$ to the full \statesp\ by the group action
$\matrixRep(\theta(\zeit))$
that defines a \emph{moving frame}.

The idea goes back to Cartan,\rf{CartanMF} and there is a rich
literature on the \mslices\ (in variety of guises) and its applications
to problems in dynamical systems theory: notable examples include the work of
Field,\rf{Field80} Krupa,\rf{Krupa90} and Ashwin and
Melbourne,\rf{AshMe97} who used slicing to prove rigorous results for
equivariant systems. Fels and Olver\rf{FelsOlver98, FelsOlver99} used the
method of moving frames to compute invariant polynomials. Haller and
Mezi\'c\rf{HaMe98} used the \mslices, under the name ``orbit projection
map'', to study three-dimensional volume-preserving flows. Our
presentation closely follows that of \refrefs{rowley_reconstruction_2000,
BeTh04}, the former of which derives the ``reconstruction equation'' for the
template fitting method of \refref{kirby_reconstructing_1992}.

\begin{figure}
\begin{center}
\includegraphics[width=0.40\textwidth]{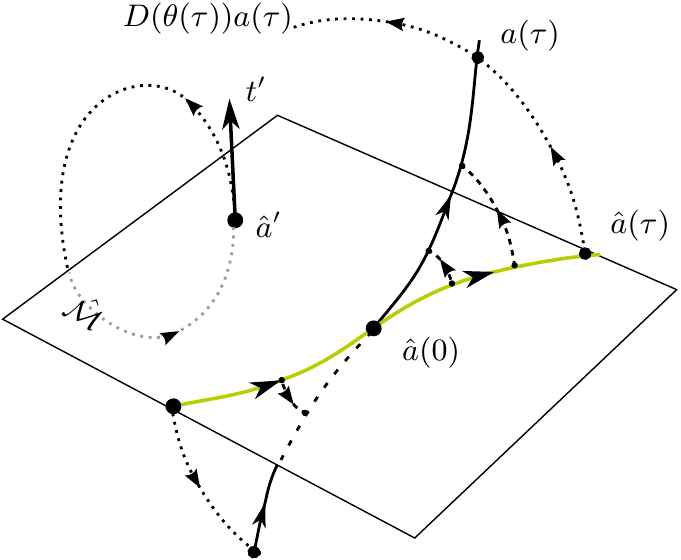}
\end{center}
\caption{\label{f-ReducTraj1}
(Color online) The \slicePlane\ \pSRed\ is a hyperplane that contains
the {\template} point $\slicep$ and is normal to its group
tangent $\sliceTan{}$. It intersects all group orbits (dotted lines) in
an open neighborhood of $\slicep$. The full \statesp\ trajectory
$\ssp(\tau)$ (solid black line) and the \reducedsp\ trajectory
$\sspRed(\zeit)$ (dashed green line) belong to the same group orbit
$\pS_{\ssp(\zeit)}$ and are equivalent up to a group rotation
$\matrixRep(\theta(\zeit))$. Adapted from \refref{DasBuch}.
}%
\end{figure}

A general definition of a \slice\ puts no restriction on its shape 
and offers no guidance on how to construct it. For Lie groups, it is 
computationally convenient to use a local, linear approximation to 
the slice (a \emph{\slicePlane}) constructed in the neighborhood of 
a point $\slicep$. (For a general discussion of how a local \slice\ 
is defined with the help of tubular neighborhood theorem, the reader 
is referred to \refrefs{Field07,Bredon72,Pal61}). This point is 
called the \emph{slice \template} and the \slicePlane\ is then 
defined as the hyperplane that contains $\slicep$ and is 
perpendicular to its group tangent $\sliceTan{} = \Lg \slicep$. The 
relationship between a \template, its \slicePlane, and 
symmetry-reduced trajectories is illustrated in \reffig{f-ReducTraj1}.

%% ReducTraj*.* - read dasbuch/book/FigSrc/inkscape/00ReadMe.txt

Reduced trajectories $\sspRed (t)$ can be obtained in two ways: by
post-processing data or by reformulating the dynamics and integrating
directly in the \slicePlane. The post-processing method (also called the
\emph{method of moving frames}\rf{FelsOlver98,OlverInv}) can be applied
to both numerical and experimental data. Here, one takes the data in the full
\statesp\ and looks for the time dependent group parameter that brings
the trajectory $\ssp(\zeit)$ onto the \slice. That is, one finds $\theta
(\zeit)$ such that $\sspRed(\zeit)=\matrixRep(-\theta (\zeit)) \ssp
(\zeit)$ satisfies the \slice\ condition:
\beq
\braket{\sspRed(\zeit) - \slicep}{\sliceTan{}} = 0
\,.
\ee{SliceCond}

In the second implementation (valid only for abelian groups),
one reformulates the dynamics as
\begin{subequations}\label{eq:so2reduced}
  \beq\label{eq:intSlice}
   \velRed(\sspRed) = \vel(\sspRed)
   -\dot{\theta}(\sspRed) \, \groupTan(\sspRed)
  \eeq
  \beq\label{eq:reconstruction}
   \dot{\theta}(\sspRed) = {\braket{\vel(\sspRed)}{\sliceTan{}}}/
            {\braket{\groupTan(\sspRed)}{\sliceTan{}}}
  \, ,
  \eeq
\end{subequations}
which can then be directly integrated to get the symmetry-reduced
trajectory $\sspRed (\zeit)$ and the reconstruction angle $\theta
(\zeit)$. In \refeq{eq:so2reduced}, $\velRed$ is the projection of the
full \statesp\ velocity \vel(\ssp) onto the \slicePlane. For a derivation,
see \refref{DasBuch}.

While early studies\rf{rowley_reconstruction_2000, rowley_reduction_2003,
BeTh04} applied the \mslices\ to a single solution at a time, studying
the nonlinear dynamics of extended systems requires symmetry reduction of
global objects, such as strange attractors and inertial manifolds. In
this spirit, \refref{SiCvi10} used the \mslices\ to
quotient the \SOn{2} symmetry from the chaotic dynamics of \cLf. They
showed that the singularity of the reconstruction equation that occurs
when the denominator in \refeq{eq:reconstruction} vanishes (e.g., when
the group tangents of the trajectory and the template are orthogonal)
causes the reduced flow to make discontinuous jumps. The set of points
$\sspRed^*$ where this occurs satisfy
\beq
\braket{\groupTan(\sspRed^*)}{\sliceTan{}} = 0
\ee{ChartBordCond}
and make up the \emph{\sliceBord} (studied in detail in \refref{FrCv11}).

Two strategies have been proposed in order to handle this problem: The first attempts to
try to identify a template such that slice singularities are not visited
by the dynamics.\rf{SiCvi10} The second uses multiple `charts' of connected
\slicePlane s,\rf{rowley_reconstruction_2000,FrCv11} switching between charts when the
dynamics approach the border of a particular chart. The latter approach was applied to \cLf\ by Cvitanovi\'{c} \etal~\rf{atlas12} and
to pipe flow by Willis, Cvitanovi\'{c}, and Avila.\rf{ACHKW11}
However, neither approach is straightforward to apply, particularly in
high-dimensional systems.

\subsection{\FFslice}
\label{sect:fFslice}

A third strategy has recently been proposed by Budanur
\etal\rf{BudCvi14}, who considered Fourier space discretizations of
partial differential equations (PDEs) with \SOn{2} symmetry. They showed
that in these cases a simple choice of \slice\ template, associated with
the first Fourier mode, results in a \slice\ in which it is highly
unlikely that generic dynamics visit the neighborhood of the singularity.
If the dynamics do occasionally come near the singularity, these close
passages can be regularized by means of a time rescaling.

Here, we shall illustrate this approach, which we call the
`\fFslice', and apply it to a model system with two modes that will be
described in \refsect{s:twoMode}.

In the discussion so far, we have not specified any constraints on the symmetry group
to be quotiented beyond the requirement that it be abelian as required for \refeq{eq:so2reduced}
to be valid. Since we are interested in spatially extended systems with
translational symmetry, and in order to keep the notation compact,
we restrict our discussion to one dimensional PDEs describing
the evolution of a field $u(x,t)$ in a periodic domain.
By expressing the solutions in terms of a Fourier series
\beq
   u(x,\zeit) = \sum\limits_{k=- \infty}^\infty u_k\left(\zeit\right) e^{i k x}, \,\,\,u_k = x_k + i y_k,
\ee{FourierSeries}
the translationally invariant PDE can be replaced by a system of coupled nonlinear
ODEs for the Fourier coefficients equivariant under the 1-parameter compact group of \SOn{2} rotations.

Truncating the expansion to $m$ modes, we write the real and imaginary
parts of the Fourier coefficients with $k \geq 1$ as the state vector
$\ssp=\cartpt{x_1, y_1, x_2, y_2,..., x_m, y_m}$. The action of the
$\SOn{2}$ group on this vector can then be expressed as a block diagonal
matrix:
\beq
   \matrixRep(\theta) = \begin{pmatrix}
                  R(\theta) & 0          & \cdots & 0 \\
                  0        & R(2 \theta) & \cdots & 0 \\
                  \vdots      & \vdots      & \ddots & \vdots \\
                  0        & 0             & \cdots & R (m \theta)
                  \end{pmatrix}
\,,
\ee{mmodeLieEl}
where
\beq
   R(n \theta) =  \begin{pmatrix}
               \cos n \theta & - \sin n \theta \\
               \sin n \theta & ~\cos n \theta
               \end{pmatrix}
\ee{rotationmatrix}
is the rotation matrix for $n$th Fourier mode.
The Lie algebra element for $\matrixRep(\theta)$ is given by
\beq
    \Lg =  \begin{pmatrix}
          0 & -1 & 0 & 0 & \cdots & 0 & 0 \\
          1 & 0 & 0 & 0 & \cdots & 0 & 0 \\
          0 & 0 & 0 & -2 & \cdots & 0 & 0 \\
          0 & 0 & 2 & 0 & \cdots & 0 & 0 \\
          \vdots & \vdots & \vdots & \vdots & \ddots & \vdots & \vdots \\
          0 & 0 & 0 & 0 & \cdots & 0 & -m \\
          0 & 0 & 0 & 0 & \cdots & m & 0
          \end{pmatrix} .
\ee{mmodeLg}

In order to construct a \slicePlane\ for such a system, we choose the
following \slice\ \template:
\beq
   \slicep = (1, 0, ..., 0) .
\ee{firstmodetemp}
The \slice\ condition \refeq{SliceCond} then constrains points on the
reduced trajectory to the hyperplane given by
\beq
   \sspRed = (\hat{x}_1, 0, \hat{x}_2, \hat{y}_2, ..., \hat{x}_m, \hat{y}_m) .
\ee{slicetemp}
As discussed earlier, group orbits should cross the \slice\ once and only
once, which we achieve by restricting the \slicePlane\ to the half-space
where $\hat{x}_1 > 0$. In general, a \slicePlane\ can be constructed by
following a similar procedure for any choice of \template. However, the power of
choosing template \refeq{firstmodetemp} becomes apparent by computing the
border \refeq{ChartBordCond} of its \slicePlane. The points on \refeq{slicetemp} lie on the
\sliceBord\ only if $\hat{x}_1 = 0$. This means that as long the dynamics
are such that the magnitude of the first mode never vanishes,
\emph{every} group orbit is guaranteed to have a unique representative
point on the \slicePlane. By symmetry, any template of the form $\slicep
=\cartpt{\hat{x}'_1, \hat{y}'_1, 0,...,0}$  would work just as well. The
\slice\ \template\ \refeq{firstmodetemp} was chosen for notational and
computational convenience.

More insight can be
gained by writing the symmetry-reduced evolution equations \refeq{eq:so2reduced}
explicitly for template \refeq{firstmodetemp}:
\begin{subequations}
\beq
\velRed ( \sspRed )  = \vel(\sspRed)
   - \frac{\dot{y}_1\left(\sspRed\right)}{\hat{x}_1} \, \groupTan(\sspRed) \, ,
\label{e-so2red1stmode}
\eeq

\beq\label{eq:reconstruction1stmode}
\dot{\theta}(\sspRed) = \frac{\dot{y}_1(\sspRed)}{\hat{x}_1}
\, .
\eeq

\end{subequations}

Since the argument $\phi_1$ of a point $(x_1,y_1)$ in the first Fourier mode plane is given by $\phi_1=\tan^{-1}\frac{y_1}{x_1}$,
its velocity is
\beq
  \dot{\phi}_1 = \frac{x_1}{r_1^2}\dot{y}_1-\frac{y_1}{r_1^2}\,\dot{x}_1\,,
\eeq
where $r_1^2=x_1^2+y_1^2$. Therefore, on the \slicePlane\ \refeq{slicetemp}, where $\hat{y}_1=0$,
\beq\label{eq:phi1}
  \dot{\theta}(\sspRed) = \dot{\phi}_1(\sspRed)\,.
\eeq
That is, for our choice of \template\ \refeq{firstmodetemp}, the
reconstruction phase coincides with the phase of the first Fourier mode.
This makes this choice of template more natural from a group-theoretic
point of view than the physically motivated templates used in
\refrefs{rowley_reconstruction_2000,BeTh04,SiCvi10,FrCv11,atlas12,ACHKW11}.

In general, additional care must be taken when the dynamics approach the
\slice\ border $\hat{x}_1 = 0$. Whenever this happens, the
near-divergence of $\velRed$ can be regularized by introducing a rescaled
time coordinate\rf{BudCvi14} such that $d\hat{\zeit} = d\zeit /
\hat{x}_1$. However, in our analysis of the \twomode\ system that we
introduce below, we omit this step since points with a vanishing first
mode are in an invariant subspace of the flow and, hence, are never
visited by the dynamics.

\subsection{Geometric interpretation of the first Fourier mode \slice}
\label{s-mframes}

\begin{figure}%[H]
\centering
 \includegraphics[width=0.45\textwidth]{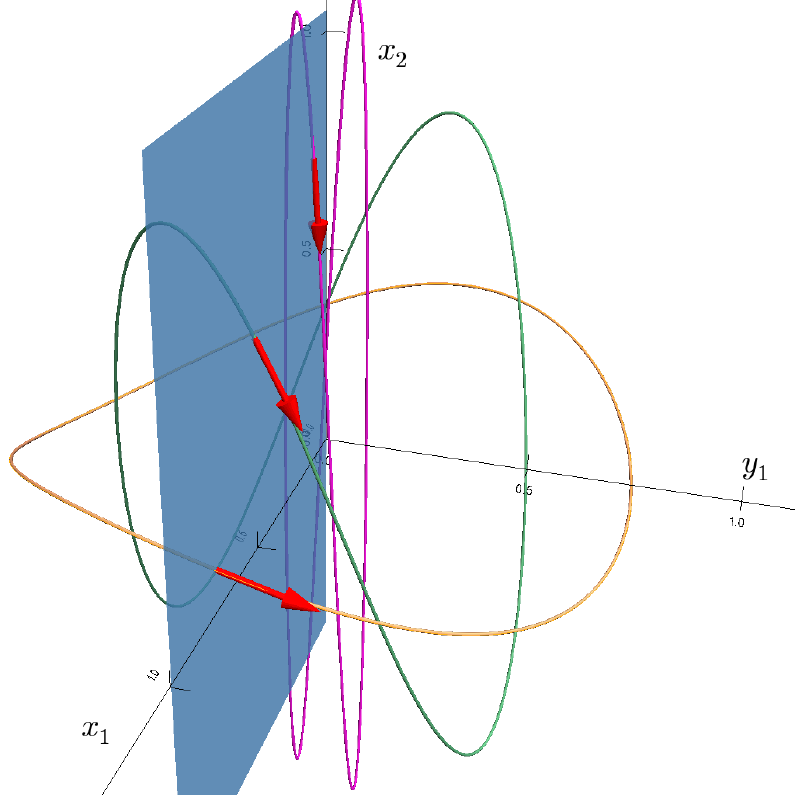}
\caption{(Color online)
$\SOn{2}$ group orbits of \statesp\ points $\cartpt{0.75, 0, 0.1, 0.1}$
(orange), $\cartpt{0.5, 0, 0.5, 0.5}$ (green)
$\cartpt{0.1, 0, 0.75, 0.75}$ (pink) and the first mode \refeq{slicetemp} \slicePlane\
(blue). The group tangents at the intersections with the
\slicePlane\ are shown as red arrows.
As the magnitude of the first Fourier mode decreases relative to the
magnitude of the second one, so does the group tangent angle to the
\slicePlane.}
\label{fig:BBgorbitsandslice}
\end{figure}

Before moving on to our analysis of the \twomode\ model, we first discuss the geometrical
interpretation of the first Fourier mode \slice. The \slice\ defined by \refeq{firstmodetemp}, along with the directional constraint
$\hat{x}_1 > 0$, fixes the phase of the first complex Fourier mode to $0$. This can also be seen
from \refeq{eq:phi1}, which shows that if the first Fourier mode \slice\ \refeq{firstmodetemp} is used as
a \template, the reconstruction phase is the same as the phase of the first
Fourier mode \refeq{eq:phi1}.

In complex representation, we can express the relationship
between Fourier modes ($\sspC_n = x_n + \ii y_n$) and their
representative points ($\sspRedC = \hat{x}_n +  \ii \hat{y}_n$) on the
\slicePlane\ by the $\Un{1}$ action:
\beq
   \sspRedC_n = e^{-\ii n \phi_1} \sspC_n \, . \ee{e-1stmodeTransform}
This relation provides another interpretation for the \sliceBord:  For
template \refeq{firstmodetemp}, the \sliceBord\ condition
\refeq{ChartBordCond} defines the \sliceBord\ as those points where $|\sspRedC_1| = |\sspC_1| = 0$.
At these points the phase of the first Fourier mode is not well-defined and hence
neither is the transformation \refeq{e-1stmodeTransform}.

This is illustrated in \reffig{fig:BBgorbitsandslice}, where the
first Fourier mode \slicePlane\ is shown along with the group orbits of points
with decreasing $|\sspC_1|$. When the magnitude of the first mode is small
relative to that of the second (pink curve), the group tangent at the representative point for
the group orbit (i.e., where the group orbit and the \slicePlane\ intersect) has a larger component
parallel to the \slicePlane. If the magnitude of the first mode was exactly
$0$, the group tangent would lie entirely on the \slicePlane , satisfying
the \sliceBord\ condition.

In \refref{PoKno05}, a polar coordinate representation of a two Fourier mode
normal form is obtained by defining the $\Group$-invariant phase: $\Phi = \phi_2 - 2 \phi_1$
and three symmetry invariant coordinates \polpt{r_1, r_2 \cos \Phi, r_2 \sin \Phi}.
One can see by direct comparison with \refeq{e-1stmodeTransform}, which
yields $\sspRedC_1 = r_1$ and $\sspRedC_2 = r_2 e^{\ii \Phi}$, that this
representation is a special case $(m=2)$, of the \slice\ defined by
\refeq{firstmodetemp}. Corresponding ODEs for the polar representation
were obtained in \refref{PoKno05} by  chain rule and substitution. Note
that the \mslices\ provides a general form \refeq{e-so2red1stmode} for symmetry
reduced time evolution.

\section{\twoMode\ $\SOn{2}$-equivariant flow}
\label{s:twoMode}

Dangelmayr,\rf{Dang86} Armbruster, Guckenheimer and Holmes,\rf{AGHO288}
Jones and Proctor,\rf{JoPro87} and Porter and Knobloch\rf{PoKno05} (for more details, see
Sect. XX.1 in Golubitsky \etal\rf{golubII}) have investigated bifurcations
in 1:2 resonance ODE normal form models to third order in the amplitudes.
Here, we use this model as a starting point from which we derive what may
be one of the simplest chaotic systems with continuous symmetry. We refer to this as the {\twomode} system:
\bea
        \dot{z}_1 &=& (\mu_1-\ii\, e_1)\,z_1+a_1\,z_1|z_1|^2
                                 +b_1\,z_1|z_2|^2+c_1\,\overline{z}_1\,z_2
        \continue
        \dot{z}_2 &=& (\mu_2-\ii\, e_2)\,{z_2}+a_2\,z_2|z_1|^2
                                 +b_2\,z_2|z_2|^2+c_2\,z_1^2 \,,
        \label{eq:DangSO2}
\eea
where $z_1$ and $z_2$ are complex and all parameters real-valued. The parameters $\{e_1,e_2\}$ break the reflectional symmetry of the $\On{2}$-equivariant normal form studied by Dangelmayr\rf{Dang86} leading to an
$\SOn{2}$-equivariant system. This complex two mode
system can be expressed as a 4-dimensional system of real-valued first order ODEs by
substituting $z_1 = x_1 + i\,y_1$, $z_2 = x_2 + i\,y_2$, so that \bea
\dot{x}_1 &=& (\mu_1 + a_1 r_1^2 + b_1 r_2^2 + c_1 x_2)x_1 + c_1 y_1 y_2 + e_1 y_1 \, ,% double checked DBE 05/22/2014
\continue
\dot{y}_1 &=& (\mu_1 + a_1 r_1^2 + b_1 r_2^2 - c_1 x_2)y_1 + c_1 x_1 y_2 - e_1 x_1 \, ,% double checked DBE 05/22/2014
\continue
\dot{x}_2 &=& (\mu_2 + a_2 r_1^2 + b_2 r_2^2)x_2 + c_2 (x_1^2 - y_1^2) + e_2 y_2 \, ,% double checked DBE 05/22/2014
\label{2mode4D}
\continue
\dot{y}_2 &=& (\mu_2 + a_2 r_1^2 + b_2 r_2^2)y_2 + 2 c_2 x_1 y_1 - e_2 x_2 \, ,% double checked DBE 05/22/2014
\continue
                  && \mbox{where } r_1^2 = x_1^2 + y_1^2\, , \quad r_2^2 = x_2^2 + y_2^2
\,.
\eea

The large number of parameters $(\mu_1,\mu_2,a_1,a_2,b_1,$ $b_2,c_1,c_2,e_1,e_2)$ 
in this system makes full exploration of the parameter space impractical. Following in the tradition of Lorenz,\rf{lorenz}
H\'enon,\rf{henon} and R\"ossler,\rf{ross} we have tried various
choices of parameters until settling on the following set of values, which we will use in all
numerical calculations presented here:
\beq
        \begin{tabular}{c c c c c c c c c c}
         $\mu_1$ & $\mu_2$ & $e_1$ & $e_2$ & $a_1$ & $a_2$ & $b_1$ & $b_2$ & $c_1$ & $c_2$ \\
        \hline
         -2.8   & 1               & 0     & 1     & -1    & -2.66 & 0     & 0     & -7.75 & 1
        \end{tabular}
        \label{eq:pars}
\eeq
This choice of parameters is far from the bifurcation values studied
by previous authors,\rf{Dang86,AGHO288,JoPro87,PoKno05} so that the
model has no physical interpretation. However, these parameters yield
chaotic dynamics, making the two-mode system a convenient minimal model
for the study of chaos in the presence of a continuous symmetry: It is a
4\dmn\ $\SOn{2}$-equivariant model, whose symmetry-reduced dynamics are
chaotic and take place on a three-dimensional manifold. For another
example of parameter values that result in chaotic dynamics, see
\refref{PoKno05}.

It can be checked by inspection that Eqs.~\refeq{eq:DangSO2} are
equivariant under the \Un{1}\ transformation
\beq
(z_1,z_2) \rightarrow   (e^{\ii {\gSpace}}z_1,e^{\ii 2{\gSpace}} z_2)
\,.
\ee{Dang86(1.1)aa}
In the real representation \refeq{2mode4D}, the $\SOn{2}$ group action
\refeq{Dang86(1.1)aa} on a state space point $\ssp$ is given $\exp\left( \theta \Lg\right)\ssp$,
where $\transp{\ssp} =\cartpt{x_1, y_1,x_2, y_2}$ and $\Lg$ is the Lie algebra
element
\beq
\Lg  \, =
\left( \begin{array}{cccc}
         0 & -1 & 0 & 0 \\
         1 & 0 & 0 & 0 \\
         0 & 0 & 0 & -2\\
         0 & 0 & 2 & 0
      \end{array} \right)
\,.
\ee{LGTwoMode}
One can easily check that the real \twomode\ system \refeq{2mode4D}
satisfies the equivariance condition \refeq{inftmInv}.

From \refeq{eq:DangSO2}, it is obvious that the \eqv\ point \((z_1,z_2)=(0,0)\)
is an invariant subspace and that $z_1=0$, $z_2 \neq 0$ is a 2\dmn\
flow-invariant subspace
\beq
  \dot{z}_1 = 0 % Double checked DBE 05/26/2014
\,,\qquad
  \dot{z}_2 = (\mu_2-\ii\, e_2 +b_2 |z_2|^2)\,{z_2} % Double checked DBE 05/26/2014
\,
\ee{eq:DangSO2spsp}
with a single circular \reqv\ of radius $r_2 = \norm{z_2} = \sqrt{-\mu_2/b_2}$ with
\phaseVel\ $\velRel=-e_2/2$. At the origin the stability matrix $\Mvar$ commutes with $\Lg$,
and so, can be block-diagonalized into two $[2\!\times\!2]$ matrices.
% According to {\bf [2012-04-27 Daniel]},
The eigenvalues of $\Mvar$ at $\cartpt{0,0,0,0}$ are $\Lyap_1 = \mu_1$ with multiplicity 2 and
$\Lyap_2 = \mu_2 \pm i e_2$. In the $\cartpt{x_1,y_1,x_2,y_2}$ coordinates, the eigenvectors for $\Lyap_1$ are $\cartpt{1,0,0,0}$ and
$\cartpt{0,1,0,0}$ and the eigenvectors for $\Lyap_2$
are $\cartpt{0,0,1,0}$ and $\cartpt{0,0,0,1}$.

In contrast, $z_2 =0$ is not, in general, a flow-invariant subspace since the dynamics
\[
  \dot{z}_1 = (\mu_1-\ii\, e_1)\,z_1+a_1\,z_1|z_1|^2
\,,\qquad
  \dot{z}_2 = c_2\,z_1^2
\,.
\]
take the flow out of the $z_2 =0$ plane.

\subsection{Invariant polynomial bases}
\label{s:invPol}

Before continuing our tutorial on the use of the method of slices using the \fFslice, we briefly discuss the symmetry reduction of the \twomode\ system using invariant polynomials. While representations of our model in terms of invariant polynomials and polar coordinates are useful for cross-checking our calculations in the full \statesp\ $\transp{\ssp} =\cartpt{x_1, x_2,y_1, y_2}$, their construction requires a bit of algebra even for this simple 4-dimensional flow. For very high\dmn\ flows, such as \KS\ and \NS\ flows, we do not know how to carry out such constructions. As discussed in \refrefs{Dang86,AGHO288,PoKno05}, for the \twomode\ system, it is easy to construct a set of four real-valued $\SOn{2}$ invariant
polynomials
\bea
u &=& {z}_1 \overline{z}_1
    \,,\quad
v = {z}_2 \overline{z}_2
    \continue
w &=& z_1^2 \overline{z}_2 + \overline{z}_1^2 {z}_2
    \,,\quad
q = (z_1^2 \overline{z}_2 - \overline{z}_1^2 {z}_2)/\ii
\,.
\label{Dang86(1.2)PK}
\eea
The polynomials $\invpt{u,v,w,q}$ are
linearly independent, but related through one syzygy,
\beq
w^2+q^2 - 4\,u^2v = 0 % Double checked syzygy is satisfied by eq Dang86(1.2)PK DBE 05/22/2014
\label{eq:syzPK}
\eeq
that confines the dynamics to a 3-dim\-ens\-ion\-al manifold $\pSRed=\pS/\SOn{2}$, which is a symmetry-invariant repre\-sent\-ati\-on of the
4-dim\-ens\-ion\-al \SOn{2} equivariant dynamics. We call this the \reducedsp. By construction, $u \geq
0$, $v \geq 0$, but $w$ and $q$ can be of either sign. That is explicit if we express $z_1$ and $z_2$ in polar coordinates ($ {z}_1 = |u|^{1/2} e^{\ii\phi_1}$, $ {z}_2 =
|v|^{1/2} e^{\ii\phi_2}$), so that $w$ and $q$ take the form
\bea
w &=& 2\,\Re(z_1^2 \overline{z}_2) = 2\,u |v|^{1/2} \cos \psi %Triple checked DBE 05/22/2014
\continue
q &=& 2\,\Im(z_1^2 \overline{z}_2) = 2\,u |v|^{1/2} \sin \psi %Triple checked DBE 05/22/2014
\,,
\label{Dang86(1.2)polar}
\eea
where $\psi = 2 \phi_1 - \phi_2$.

The dynamical equations for $\invpt{u,v,w,q}$ follow from the chain rule,
which yields
\bea
  \dot{u} &=& \overline{z}_1 \dot{z}_1 + {z}_1 \dot{\overline{z}}_1 % Triple checked DBE 05/22/2014
\,,\qquad
  \dot{v} = \overline{z}_2 \dot{z}_2 + {z}_2 \dot{\overline{z}}_2 % Triple checked DBE 05/22/2014
\continue
  \dot{w} &=& 2 \,\overline{z}_2 {z}_1 \dot{z}_1 % Triple checked DBE 05/22/2014
           + 2\,{z}_2 \overline{z}_1 \dot{\overline{z}}_1
           + {z}_1^2 \dot{\overline{z}}_2
           + \overline{z}_1^2 \dot{z}_2
\continue
  \dot{q} &=&  (2\,\overline{z}_2 {z}_1 \dot{z}_1 % Triple checked DBE 05/22/2014
           - 2\,{z}_2 \overline{z}_1 \dot{\overline{z}}_1
           + {z}_1^2 \dot{\overline{z}}_2
           - \overline{z}_1^2 \dot{z}_2
           )/\ii
\label{PKinvEqs}
\eea
Substituting \refeq{eq:DangSO2} into \refeq{PKinvEqs}, we obtain a set
of four $\SOn{2}$-invariant equations,
\bea
  \dot{u} &=& 2\,\mu_1\,u+2\,a_1\,u^2+2\,b_1\,u\,v+c_1\,w % Triple checked DBE 05/22/2014
\continue
  \dot{v} &=& 2\,\mu_2\,v+2\,a_2\,u\,v+2\,b_2\,v^2+c_2\,w % Triple checked DBE 05/22/2014
\continue
  \dot{w} &=& (2\,\mu_1+\mu_2)\,w+(2a_1+a_2)\,u\,w+(2b_1+b_2)\,v\,w % Triple checked DBE 05/22/2014
\ceq
             +\, 4c_1\,u\,v + 2c_2\,u^2 +(2e_1 - e_2)\,q
\label{PKinvEqs1}\\
  \dot{q} &=& (2\mu_1+\mu_2)\,q+(2a_1+a_2)\,u\,q
\ceq
             +(2b_1+b_2)\,v\,q
             -(2e_1-e_2)\,w % Triple checked DBE 05/22/2014
\,.
\nnu
\eea
Note that the $\On{2}$-symmetry breaking parameters
 $\{e_1,e_2\}$ of the
Dangelmayr normal form system\rf{Dang86} appear only in the
relative phase combination $(2e_1-e_2)$, so one of the two can be set to zero without loss of generality. This consideration motivated our choice of $e_1 = 0$ in \refeq{eq:pars}.
%[2012-07-31 Evangelos]
Using the syzygy \refeq{eq:syzPK}, we can
eliminate $q$ from \refeq{PKinvEqs1} to get
\bea
  \dot{u} &=& 2\,\mu_1\,u+2\,a_1\,u^2+2\,b_1\,u\,v+c_1\,w \nonumber % Triple checked DBE 05/22/2014
\\
  \dot{v} &=& 2\,\mu_2\,v+2\,a_2\,u\,v+2\,b_2\,v^2+c_2\,w \label{PKinvEqs1syz}  % Triple checked DBE 05/22/2014
\\
  \dot{w} &=& (2\,\mu_1+\mu_2)\,w+(2a_1+a_2)\,u\,w+(2b_1+b_2)\,v\,w % Triple checked DBE 05/22/2014
\ceq
             +\, 4c_1\,u\,v + 2c_2\,u^2 +(2e_1 - e_2)(4u^2v-w^2)^{1/2}\,
  \nonumber
\eea
This invariant basis can be used either to investigate the dynamics directly or
to visualize solutions\rf{GL-Gil07b} computed in the full equivariant basis \refeq{eq:DangSO2}.

\subsection{\Eqva\ of the symmetry-reduced dynamics}
\label{s:eqva}

The first step in elucidating the geometry of attracting sets is the
determination of their \eqva. We shall now show that the problem of
determining the \eqva\ of the symmetry-reduced \twomode\
\refeq{PKinvEqs1} system $\invpt{u^*,v^*,w^*,q^*}$ can be reduced to
finding the real roots of a multinomial expression. First, we define
\beq
A_1= \mu_1+a_1\,u+b_1\,v
    \,,\qquad
A_2 = \mu_2+a_2\,u+b_2\,v
\ee{PKinvEqs2a}
and rewrite \refeq{PKinvEqs1} as
%     \newpage
\bea
  0  &=&  2\,A_1\,u +c_1\,w % Double checked DBE 05/24/2014
    \,,\qquad
  0  =  2\,A_2\,v +c_2\,w % Double checked DBE 05/24/2014
\continue
  0  &=& (2\,A_1+ A_2)\,w
          +2\,\left(c_2\,u+2\,c_1\,v\right)\,u % Double checked DBE 05/24/2014
          \ceq
                  + (2e_1-e_2)\,q
\label{PKinvEqs3}\\
  0  &=& (2\,A_1+ A_2)\,q - (2e_1-e_2)\,\,w % Double checked DBE 05/24/2014
\nnu
\eea
We already know that $\invpt{0,0,0,0}$ and $\invpt{0,-\mu_2/b_2,0,0}$
are the only roots in the $u=0$ and $v=0$ subspaces, so we are
looking only for the $u>0$, $v>0$, $w,q \in \reals$ solutions; there
could be non-generic roots with either $w=0$ or $q=0$, but not both
simultaneously, since the syzygy \refeq{eq:syzPK} precludes that. Either
$w$ or $q$ can be eliminated by obtaining the following relations from
\refeq{PKinvEqs3}:
\bea
        w  &=& - \frac{2\,u}{c_1}\,A_1 = - \frac{2\,v}{c_2}\,A_2 % Double checked DBE 05/24/2014
        \continue
        q &=& \frac{2(-2e_1+\,e_2)\,u\,v}{c_2\,u+2\,c_1\,v} . % Having issues with this DBE 05/24/2014... potentially drops a w = 0 root.
        \label{PKinvEqs4}
\eea
Substituting \refeq{PKinvEqs4} into \refeq{PKinvEqs3} we get two bivariate
polynomials whose roots are the \eqva\ of the system \refeq{PKinvEqs1}:
\bea
        f(u,v) &=& c_2\,u\,A_1 - c_1\,v\,A_2 = 0 \,,\qquad  \nonumber
        \\
        g(u,v) &=&
 \left(4\,A_1^2 u^2 - 4\,c_1^2\,u^2 v\right)\left(c_2\,u+2\,c_1\,v\right)^2 \label{PKinvEqs5} %Double checked DB 04-30-2012
        \ceq
        +\,4\,c_1^2\,(-2e_1+e_2)^2\,u^2\,v^2 = 0
\,.
\eea
We divide the common multiplier $u^2$ from the second equation and by
doing so, eliminate one of the two roots at the origin, as well as the
$\invpt{0,-\mu_2/b_2,0,0}$ root within the invariant subspace
\refeq{eq:DangSO2spsp}. Furthermore, we scale the parameters and
variables as
$\tilde{u} = c_2\,u$,
$\tilde{v} = c_1\,v$,
$\tilde{a_1} = a_1/c_2$,
$\tilde{b_1} = b_1/c_1$,
$\tilde{a_2} = a_2/c_2$,
$\tilde{b_2} = b_2/c_1$
to get
\bea
\tilde{f}(\tilde{u},\tilde{v}) &=&
  \tilde{u}\,\tilde{A}_1 - \tilde{v}\,\tilde{A}_2 = 0 %Double checked DB 04-30-2012
\,, \label{PKinvEqs5a}
\\
\tilde{g}(\tilde{u},\tilde{v}) &=&  %Double checked DB 04-30-2012
 \left(\tilde{A}_1^2
 - c_1\,\tilde{v}\right)
 \left(\tilde{u}+2\,\tilde{v}\right)^2
 +e_2^2\,\tilde{v}^2 = 0
\,, \label{PKinvEqs5b}
\eea

\noindent where $\tilde{A}_1 = \mu_1+\tilde{a_1}\,\tilde{u}+\tilde{b_1}\,\tilde{v}$ and
$\tilde{A}_2 = \mu_2+\tilde{a_2}\,\tilde{u}+\tilde{b_2}\,\tilde{v}$.

Solving coupled bivariate polynomials such as \refeq{PKinvEqs5a} and \refeq{PKinvEqs5b}, is not, in general, 
a trivial task. However, for the choice of parameters given by \refeq{eq:pars}, Eq.~\refeq{PKinvEqs5a} yields
$\tilde{v} = (\mu_1 + \tilde{a}_1 \tilde{u})/(\mu_2 + \tilde{a}_2
\tilde{u})$. Substituting this into \refeq{PKinvEqs5b} makes it a fourth order polynomial in $u$,
which we can solve. Only the non-negative, real roots of this polynomial correspond to \reqva\ in the \twomode\
\statesp\ since $u$ and $v$ are the squares of first and second mode amplitudes,
respectively. Two roots satisfy this condition, the \eqv\ at the origin
\beq
        \invpol_{\EQV{}} = \invpt{0,0,0,0}\,, %\qquad \mbox{(double)}
\ee{eq:origin}
and the \reqv\
\beq
        \invpol_{\REQV{}{}} = \invpt{0.193569,0.154131,-0.149539,-0.027178}\,.
\ee{eq:reqv}
Note that by setting $b_2 = 0$, we send the \reqv\ at
$\invpt{0,-\mu_2/b_2,0,0}$ to infinity. Thus, \refeq{eq:reqv} is the
only \reqv\ of the \twomode\ system for our choice of parameters. While
this is an \eqv\ in the invariant polynomial basis, in the
\SOn{2}-equivariant, real-valued \statesp\ this is a 1\dmn\ \reqv\ group orbit.
The point on this orbit that lies in first Fourier mode slice is
(see \refFig{fig:2modes-ssp}\,(c)):
\beq
  \left(x_1, y_1, x_2, y_2\right) = \left(0.439966, 0, -0.386267, 0.070204\right)
\,.
  \label{e-req}
\eeq
We computed the linear stability eigenvalues and eigenvectors of this \reqv
, by analyzing the \stabmat\ within the first Fourier mode slice
$\MvarRed_{ij} (\sspRed) = \partial \velRed_i / \partial \sspRed_j |_{\sspRed}$, 
resulting in linear stability eigenvalues %\bea
%       \lambda_{1,2} &=& 0.05073 \pm \ii 2.4527, \continue
%       \lambda_3 &=& -5.5055, \quad \lambda_4 = 0 \, .
%\eea
\beq
        \lambda_{1,2} = 0.05073 \pm \ii \, 2.4527, \quad
        \lambda_3 = -5.5055, \quad \lambda_4 = 0 \, .
\eeq
The $0$ eigenvalue corresponds to the direction outside the slice. We expect
this since the reduced trajectory evolution equation \refeq{eq:intSlice} keeps the
solution within the slice. The imaginary part of the expanding complex pair sets
the `winding time' in the neighborhood of the equilibrium to
$T_w = 2 \pi / \Im(\lambda_1) = 2.5617$. The large magnitude of the
contracting eigenvalue $\lambda_3$ yields a very thin attractor in the
reduced \statesp, thus, when looked at on a planar Poincar\'{e} section,
the \twomode\ flow is almost one dimensional, as shown in Figs. \ref{fig:psectandretmap}(a) and \ref{fig:psectandretmap}(b).

\begin{figure*}%[H]
\centering
\includegraphics[height=0.22\textwidth]{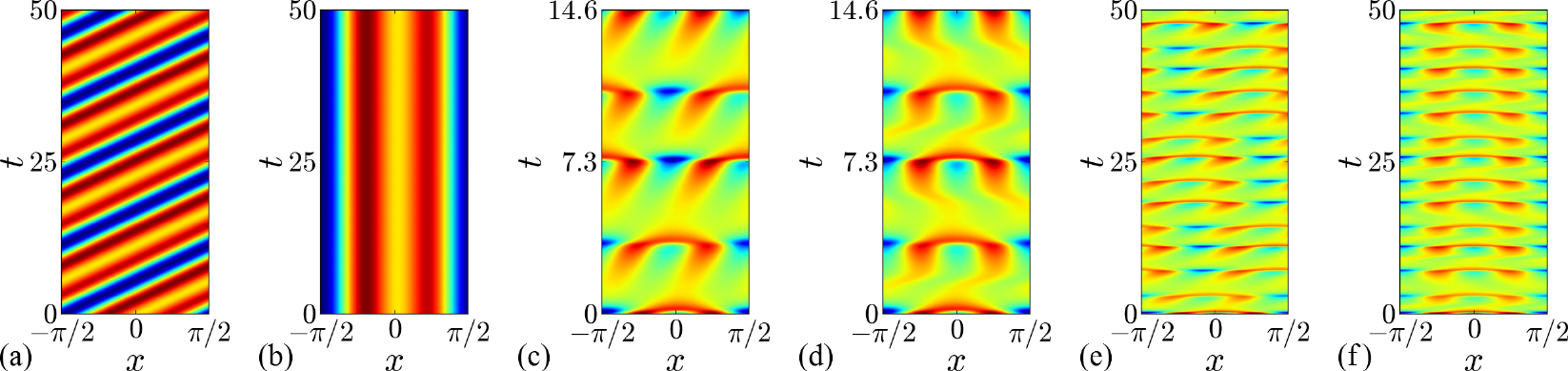}
\caption{(Color online)
The \reqv\ \REQV{}{} in
 (a) the system's configuration space becomes an \eqv\ in
 (b) the symmetry-reduced configuration space.
Two cycles of the \rpo\ \cycle{01} in the
 (c) the symmetry-equivariant configuration space become a \po\ in
 (d) the symmetry-reduced configuration space.
 (e) A typical ergodic trajectory of the \twomode\ system
in the system's configuration space,
 (f) in the symmetry-reduced configuration space.
The color scale used in each figure is different to enhance contrast.
}
\label{fig:2modes-conf}
\end{figure*}

\begin{figure*}%[H]
\centering
\includegraphics[height=0.25\textwidth]{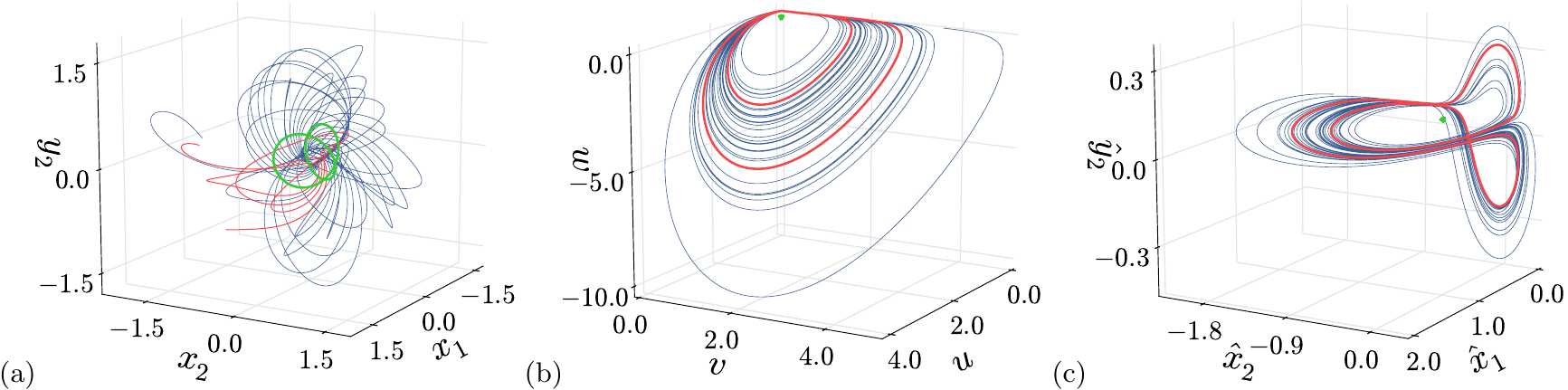}
\caption{(Color online)
The trajectories as in Figs. \ref{fig:2modes-conf}\,(a,c,e) are
colored green, red and blue respectively,
        (a) in a 3D projection of the 4\dmn\ \statesp ,
        (b) in a terms of 3 invariant polynomials,
        (c) in the 3\dmn\  first Fourier mode \slicePlane.
Note that in the symmetry reduced representations (b and c), the \reqv\ \REQV{}{}
is reduced to an \eqv , the green point; and the periodic orbit \cycle{01} (red) closes onto
itself after one repeat. In contrast to the invariant polynomial representation (b),
in the first Fourier mode \slicePlane (c), the qualitative difference between shifts
by $\approx \pi$ and $\approx-\pi$ in near passages to the {\sliceBord} is very clear,
and it leads to the unimodal Poincar\'e return map of \reffig{fig:psectandretmap}.
}
\label{fig:2modes-ssp}
\end{figure*}

\subsection{No chaos when the reflection symmetry is restored}
\label{s:dfsafs}

Before finishing our discussion of invariant polynomials, we make an
important observation regarding the case when both of the reflection symmetry breaking
parameters, $e_{1}$ and $e_2$ are set to $0$. In this case, $\sspC_{1,2} \rightarrow \bar{\sspC}_{1,2}$
symmetry is restored and the evolution equations for $u$, $v$, and $w$ in \refeq{PKinvEqs1} become
independent of $q$. Furthermore, the time evolution equation for $q$ becomes linear in $q$ itself, so that
it can be expressed as
\beq
    \dot{q} = \xi (u, v) q \,.
\ee{e-qlinearq}
Hence, the time evolution of $q$ can be written as
\beq
    q(\zeit) =  e^{\int_0^\zeit d \zeit' \xi (u(\zeit'), v(\zeit'))} q(0) \, .
\ee{e-qO2solq}
If we assume that the flow is bounded, then we can also assume that a long time
average of $\xi$ exists. The sign of this average determines the long term
behavior of $q(\zeit)$; it will either diverge or vanish depending on the sign of
$\langle \xi \rangle$ being positive or negative respectively. The former case
leads to a contradiction: If $q(\zeit)$ diverges, the symmetry-invariant flow cannot
be bounded since the syzygy \refeq{eq:syzPK} must be satisfied at all times. If
$q(t)$ vanishes, there are three invariant polynomials left, which are still
related to each other by the syzygy. Thus, the flow is confined
to a two dimensional manifold and cannot exhibit chaos.
We must stress that this is a special result that holds for the two-mode
normal form with terms up to third order.

\subsection{Visualizing \twomode\ dynamics}
\label{s:visual}

We now present visualizations of the dynamics of the \twomode\ system in
four different representations: as 3D projections of the four-dimensional
real-valued \statesp, as 3D projections in the invariant polynomial
basis, as dynamics in the 3D \slicePlane, and as two-dimensional
spacetime diagrams of the color-coded field
$u(\conf,\zeit),$ which is defined as follows:
\[
        u(\conf, \tau) = \sum_{k=-2}^{2} \sspC_k(\zeit) \, e^{i k \conf}
\,,
\]
where $\sspC_{-k} = \bar{\sspC}_k \,, \;        \sspC_0 = 0$ ,  and $\conf
\in [- \pi, \pi]$. We can also define the symmetry reduced configuration
space representation as the inverse Fourier transform of the symmetry
reduced Fourier modes:
\[
        \hat{u}(\conf, \tau) = \sum_{k=-2}^{2} \sspRedC_k(\zeit) e^{i k \conf}
\,,
\]
where $\sspRedC_{-k} = \bar{\sspRedC}_k$ \,, \;         $\sspRedC_0 = 0$ \;
and $\conf \in [- \pi, \pi]$. Figures \ref{fig:2modes-conf}(a) and \ref{fig:2modes-conf}(b) 
show the sole \reqv\ \REQV{}{} of the \twomode\ system in the symmetry-equivariant
and symmetry-reduced configuration spaces, respectively. After
symmetry reduction, the \reqv\ becomes an \eqv.
Figures \ref{fig:2modes-conf}(c) and \ref{fig:2modes-conf}(d) show the \rpo\ \cycle{01} again
respectively in the symmetry-equivariant and symmetry-reduced
configuration space representations. Similar to the \reqv, the \rpo\
becomes a \po\ after symmetry reduction. Finally,
Figs. \ref{fig:2modes-conf}(e) and \ref{fig:2modes-conf}(f) show a typical ergodic trajectory of the
\twomode\ system in symmetry-equivariant and symmetry-reduced
configuration space representations. Note that in each case, symmetry
reduction cancels the `drifts' along the symmetry ($x$) direction.

As can be seen clearly in \reffig{fig:2modes-ssp}\,(a), these drifts show up in
the Fourier mode representation as $\SOn{2}$ rotations. The \reqv\ \REQV{}{}
traces its \SOn{2} group orbit (green curve in \reffig{fig:2modes-ssp}\,(a))
as it drifts in the configuration space. The
\rpo\ \cycle{01}\,(red) and the ergodic trajectory (blue) rotate
in the same fashion as they evolve. Figures \ref{fig:2modes-ssp}(b) and \ref{fig:2modes-ssp}(c)
show a three dimensional projection onto the invariant polynomial basis and the 3\dmn\
trajectory on the \slicePlane\ for the same orbits. In both figures, the \reqv\ is reduced
to an \eqv\ and the \rpo\ is reduced to a \po.

\section{\Po s}
\label{s:numerics}

\begin{figure}
\centering
  \includegraphics[width=0.48\textwidth]{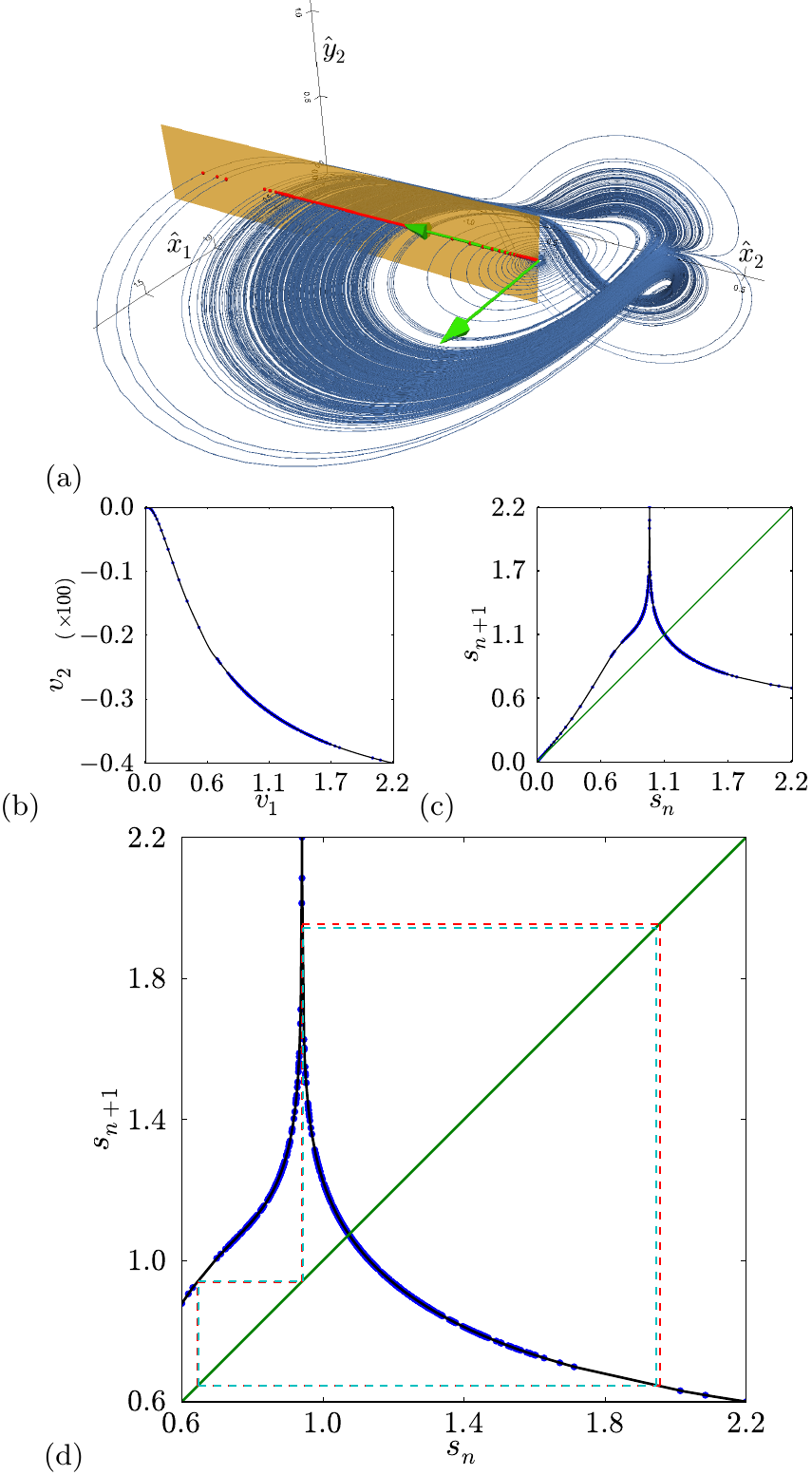} 
\caption{(Color online)
         (a) A Symmetry-reduced ergodic trajectory within the slice hyperplane 
             (blue). Green arrows indicate the real and imaginary parts of the 
             complex eigenvectors $v_u$ that span the unstable manifold of 
             \REQV{}{}. The Poincar\'e section, which contains \REQV{}{} and 
             is spanned by $\Im[v_u]$ and $\hat{y}_2$, is visualized as a 
             transparent plane. Points where the flow crosses the section are marked 
             in red.
                 (b) A closer look at the Poincar\'e section shows that the attractor is very thin.
                         Note that the vertical axis, which corresponds to the direction 
                         parallel to $\hat{y}_2$ is magnified by $100$. All (blue) points 
                         are located relative to the \REQV{}{}, which is at the origin.
                         The black curve is a cubic spline interpolation of the data.                   
                 (c) By measuring arclengths $s$ along the interpolation curve, a return map 
                 of the Poincar\'e section can be constructed. 
                         Note that once the flow exits the neighborhood of the \REQV{}{} 
                         ($s < 0.6$) it stays on the attractor and never comes back. Thus the data 
                         up to this point is transient.
                 (d) The return map without the transient points framed by orbit of the 
                     critical point. Dashed lines show the 3-cycles \cycle{001} (red) and 
                     \cycle{011} (cyan).}
\label{fig:psectandretmap}
\end{figure}

The simple structure of the symmetry-reduced dynamics allows us to
determine the \rpo s of the \twomode\ system by means of a Poincar\'e
section and a return map. We illustrate this procedure in
\reffig{fig:psectandretmap}. Starting with an initial point close to the
\REQV{}{}, we compute a long, symmetry-reduced ergodic trajectory by integrating
\refeq{e-so2red1stmode} and record where it crosses the Poincar\'e section, which we
define as the plane that contains \REQV{}{} and is spanned the imaginary part of its unstable stability
eigenvector and $\hat{y}_2$.
We then project these points onto a basis $(v_1, v_2)$, which
spans the Poincar\'e section and fit cubic splines to the data as shown in \reffig{fig:psectandretmap}\,(b).
This allows us to construct a return map along this curve, which can be expressed in terms of the distance $s$ from \REQV{}{}
as measured by the arc length along the cubic spline fit. The resulting map, which is shown in
\reffig{fig:psectandretmap}\,(c), is unimodal with a sharp cusp located at its critical point.
Note that the region $s \in (0, 0.6)$ corresponds to
the neighborhood of the \reqv\  and is only visited transiently. Once the dynamics fall onto the chaotic
attractor, this region is never visited again. Removing this region from the return map, we
obtain the return map shown in \reffig{fig:psectandretmap}\,(d), which we can then use to determine the
accessible \rpo s  with their respective binary symbol sequences.

The unimodal return map of \reffig{fig:psectandretmap} diverges around
$s \approx 0.98$ and this neighborhood is visited very rarely by the flow. We
took the furthest point that is visited by the ergodic flow, $s_C=0.98102264$
as the critical point of this map and coded points to the left and right hand sides of this
point as `0' and `1', respectively, and constructed a binary symbolic dynamics.
Accessible periodic orbits are then those with the topological coordinates
less than that of this critical point. We skip the technical details
regarding symbolic dynamics and kneading theory in this tutorial since
there is a rich literature on these topics and we do not employ any novel
symbolic dynamics technique here. For a pedagogical introduction to the
subject, we refer the reader to \refrefs{devnmap, DasBuch}.

We are now going to summarize the procedure of locating \rpo s in the \statesp :
Suppose the binary itinerary
$\cycle{I_0 I_1 \dots\ I_{n-1}}, \mbox{where,}\, I_j = 0,1$
corresponds to an admissible `n-cycle', a \rpo\ that intersects our Poincar\'e
section n-times. We first find arc-lengths $\{s_0,\,s_1,\,\dots\,s_n\}$ that
constitute this cycle on the return map \reffig{fig:psectandretmap}\,(d). We
then find corresponding reduced \statesp\ points
$\{\sspRed_0,\,\sspRed_1,\,\dots\, \sspRed_{n-1}\}$. Finally, we integrate the
reduced flow and the phase \refeq{eq:so2reduced} starting from each point $\sspRed_j$ until it returns to the Poincar\'e
section, and divide this trajectory into $N$ small pieces. As a result, we obtain
$n \times N$ \statesp\ points, durations and phase shifts
$\{\ssp_i^{(0)}\,,\,\zeit_i^{(0)}\,,\,\theta_i^{(0)}\}$ where
$i=1,\,2,\,\dots\,n \times N$ , which we feed into the multiple shooting Newton
solver (see \refappe{s:newton}) to precisely determine the \rpo , its period
and the associated phase shift. After finding $n \times N$ \statesp\ points
($\ssp_i$), flight times ($\zeit_i$), and phase shifts ($\theta_i$) associated
with the $n$ cycle, we can compute the stability of the orbit. We do this by computing 
the flow Jacobian associated with each segment of the orbit $\jMps^{\zeit_i}(\ssp_i)$, so that 
the Jacobian associated with the \rpo\ is then
\bea
    \jMpsRed &=&
    \matrixRep(- \theta_{n \times N} ) \jMps^{\zeit_{n \times N}} (\ssp_{n \times N})
    \dots \, \continue
    && \matrixRep(- \theta_2 ) \jMps^{\zeit_2} (\ssp_2)
       \matrixRep(- \theta_1 ) \jMps^{\zeit_1} (\ssp_1) \, .
    \label{e-MultiShootJacobian}
\eea
This construction \refeq{e-MultiShootJacobian} of the Jacobian is equivalent to our
definition in \refeq{e-rpoJacobian}, since the group action $\LieEl$ and the flow
Jacobian $\jMps$ are both multiplicative and commute with each other as a
consequence of $\LieEl$-equivariance of the flow. The form
\refeq{e-MultiShootJacobian} is essential in determining its eigenvalues
(Floquet multipliers) precisely, since it allows us to use periodic Schur
decomposition, as described in \refAppe{s:schur}.

We found the admissible cycles of the
\twomode\ system up to the topological length 12. We listed binary itineraries
of shortest $7$ \rpo s (with topological lengths up to 5), along with their
periods, phase shifts, Floquet multipliers, and Floquet exponents in
\refTab{t-rpofirst10}. In \reffig{f-2modesrpofirst4} we show shortest $4$ of
the \rpo s of the \twomode\ system within the first Fourier mode \slicePlane .
As seen from \reffig{f-2modesrpofirst4}, trajectories of \cycle{001} (red) and
\cycle{011} (cyan) almost overlap in a large region of the \statesp . This
behavior is also manifested in the return map of
\reffig{fig:psectandretmap}\,{d), where we have shown cycles \cycle{001} and
\cycle{011} with red and cyan respectively. This is a general property of the
\twomode\ cycles with odd topological lengths: They come in pairs with almost
equal leading (largest) Floquet exponents, see \reffig{f-2modes-lambdaDist}.
Floquet exponents ($\Lyap_j$) characterize the rate of expansion/contraction
of nearby perturbations to the \rpo s and are related to Floquet multipliers
($\ExpaEig_j$) by
\beq
    \Lyap_{\rpprime,j} = \frac{1}{\period{\rpprime}}
                         \ln | \ExpaEig_{\rpprime,j} |
                         \, , \quad j=1,2,\dots,d \, ,
\eeq
where the subscript $\rpprime$ associates $\Lyap_{\rpprime,j}$ and $\ExpaEig_j$
with the `prime \rpo' $p$ and its period
$\period{\rpprime}$. Having computed periods, phase shifts,
and Floquet multipliers of \rpo s, we are now ready to calculate dynamical
averages and other statistical moments of observables using \cycForm s.

\begin{table}
        \caption{Itinerary, period ($T$), phase shift ($\theta$), 
                         Floquet multiplier ($\ExpaEig$), and Floquet exponent
                         ($\Lyap$) of the found \twomode\ \rpo s with topological
                         lengths up to $n = 5$, more (up to $n=12$) available 
                         upon request.}
        \begin{tabular}{c|c|c|c|c}
        Itinerary & $T$ & $\theta$ & $\ExpaEig$ & $\Lyap$ \\ 
        \hline
        1 & 3.64151221 & 0.08096967 & -1.48372354 &0.10834917 \\ 
        01 & 7.34594158 & -2.94647181 & -2.00054831 &0.09439516 \\ 
        001 & 11.07967801 & -5.64504385 & -55.77844510 &0.36295166 \\ 
        011 & 11.07958924 & -2.50675871 & 54.16250810 &0.36030117 \\ 
        0111 & 14.67951823 & -2.74691247 & -4.55966852 &0.10335829 \\ 
        01011 & 18.39155417 & -5.61529803 & -30.00633820 &0.18494406 \\ 
        01111 & 18.38741006 & -2.48213868 & 28.41893870 &0.18202976 \\ 
        \end{tabular}
        \label{t-rpofirst10}
\end{table}

\begin{figure}%[H]
\centering
 \includegraphics[width=0.45\textwidth]{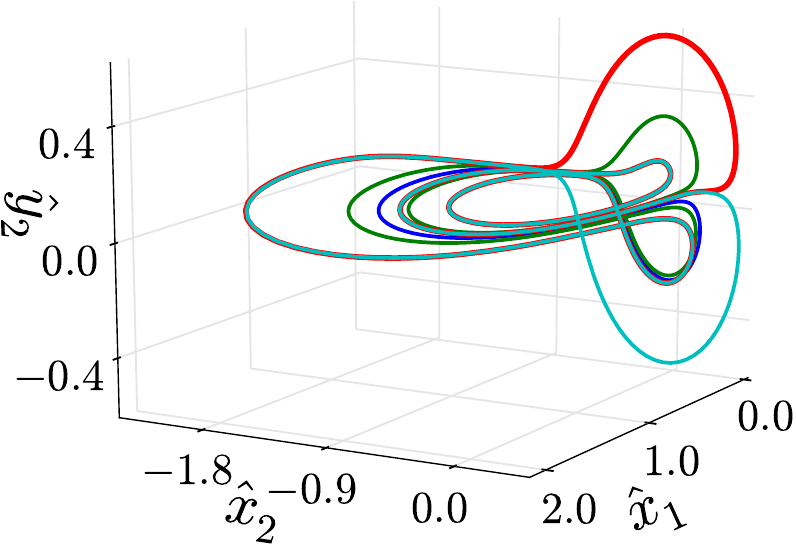}
\caption{(Color online)
Shortest four \rpo s of the \twomode\ system: \cycle{1} (dark blue),
\cycle{01} (green), \cycle{001} (red), \cycle{011} (cyan). Note that \rpo
s \cycle{001} and \cycle{011} almost overlap everywhere except $\hat{x}_1
\approx 0$ .}
\label{f-2modesrpofirst4}
\end{figure}

\begin{figure}%[H]
\centering
 \includegraphics[width=0.45\textwidth]{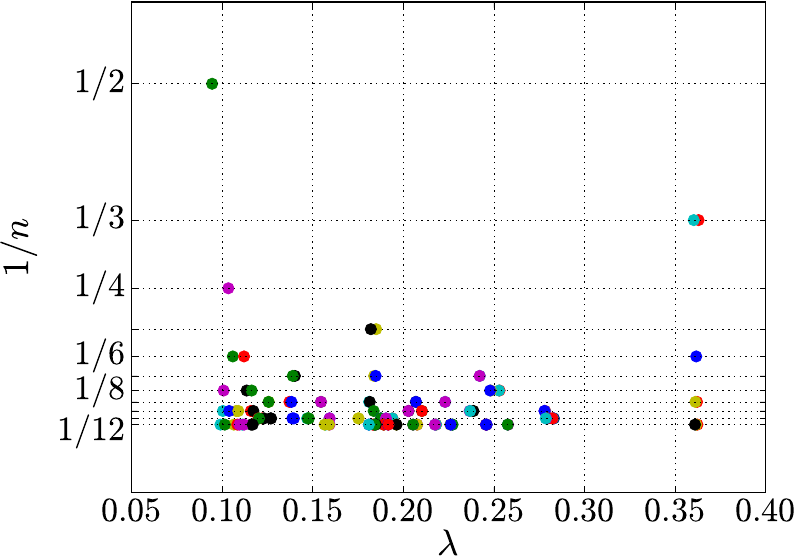}
\caption{(Color online)
        Distribution of the expanding Floquet exponents of all \twomode\ cycles with
         topological lengths $n$ from $2$ to $12$.}
\label{f-2modes-lambdaDist}
\end{figure}

\section{Cycle Averages}
\label{s:DynAvers}

\renewcommand{\zeit}{\ensuremath{t}}  %time variable

So far, we have explained how to find the \rpo s of the \twomode\ system
and compute their stability. However, we have not yet
said anything about what to do with these numbers. We begin this section with
an overview of the main results of the periodic orbit theory. Our review starts
by recapitulating the presentation of \refref{DasBuch}, but then, in
\refsect{s-ContFac}, explains how
the theory is modified in the presence of continuous symmetries.\rf{Cvi07}
In \refsect{s-CycExp}, we present cycle expansions and
explain how to approximate the Poincar\'e section in
\reffig{fig:psectandretmap}\,(d), in order to obtain a better convergence of
the spectral determinants. The numerical results are discussed in
\refsect{s-NumResults}.

\subsection{Classical trace formula}

Consider the {\evOper} $\Lop^\zeit$, the action of which evolves a
weighted density $\rho(\ssp,\zeit)$ in the \statesp,
\bea
    \rho(\ssp',\zeit) &=& [\Lop^\zeit \rho ] (\ssp')
    = \int\!d \ssp \, \Lop^\zeit (\ssp', \ssp) \, \rho(\ssp,0)
    \continue
    \Lop^\zeit (\ssp', \ssp)
    &=&
    \delta (\ssp' - \flow{\zeit}{\ssp})\,
        e^{\beta \Obser^\zeit(\ssp)}
\,,
\label{e-EvOper}
\eea
where $\beta$ is an auxiliary variable and $\Obser^\zeit (\ssp)$ is the
integrated value of an observable $\obser (\ssp)$ along the trajectory
$\ssp(\zeit)=\flow{\zeit}{\ssp}$,
\beq
    \Obser^\zeit (\ssp ) = \int_0^{\zeit} d \zeit'
                              \obser(\flow{\zeit'}{\ssp})
\,.
\eeq
When $\beta = 0$, the \evOper\ \refeq{e-EvOper} evolves the initial density of
\statesp\ points to its new form after time $\zeit$; this form of the
evolution operator is known as the {\FPoper}. The
multiplicative weight $\exp(\beta \Obser^\zeit(\ssp))$
will enable us to compute the value of the observable $\obser$ averaged over
the natural measure.

As the integrated observable $\Obser^\zeit (\ssp)$, additive along the
trajectory, is exponentiated in \refeq{e-EvOper}, the \evOper\
is multiplicative along the trajectory:
\beq
        \Lop^{\zeit_1 + \zeit_2} (\ssp',\ssp) =
    \int d\ssp'' \Lop^{\zeit_2} (\ssp', \ssp'')
                   \Lop^{\zeit_1} (\ssp'', \ssp) \, .
        \label{eq-SemiGroupKernel}
\eeq
This semigroup property allows us to define the {\evOper} as the formal
exponential of its infinitesimal generator \Aop:
\beq
        \Lop^\zeit = e^{\Aop \zeit} \, .
        \label{eq-EvOpExp}
\eeq
Let $\rho_{\beta} (\ssp)$ be the eigenfunction of
\refeq{e-EvOper} corresponding to the leading eigenvalue of $\Aop$ (\ie, the one
with the largest real part) for a given $\beta$,
\beq
    \left[ \Lop^\zeit \rho_{\beta} \right] (\ssp)
    =
    e^{\zeit s(\beta )} \rho_{\beta}(\ssp)
    \, .
    \label{eq-EigenvalueRel}
\eeq
If the system under study is
ergodic, then an invariant `natural measure' $\rho_0(\ssp)$ with
eigenvalue $s(0) = 0$ exists, and the long time average of an observable is
then its \statesp\ average over the natural measure:
\beq
    \langle \obser \rangle = \int d \ssp \, \obser(\ssp) \rho_0 (\ssp) \, .
    \label{e-obserAvg}
\eeq
By evaluating the action of the {\evOper} \refeq{e-EvOper} for
infinitesimal times, one finds that the
long-time averages of
observables, as well as of their higher moments, are given by
derivatives of $s(\beta)$:
\bea
    \langle \obser \rangle &=&
    \lim_{\zeit \rightarrow \infty} \frac{1}{\zeit} \langle \Obser^\zeit \rangle =
        \left. \frac{\partial s(\beta) }{\partial \beta} \right|_{\beta = 0}
        \, , \continue
        \Delta \; &=&
        \lim_{\zeit \rightarrow \infty} \frac{1}{\zeit}
        \langle (\Obser^{\zeit})^2 - \langle \Obser^\zeit \rangle^2 \rangle =
        \left. \frac{\partial^2 s(\beta) }{\partial \beta^2} \right|_{\beta = 0}
    \, ,
    \label{eq-moments}
    \\
    \, &\vdots & \nonumber
\eea
For example, if the observable \obser\ is a velocity
$\dot{\ssp}=\vel(\ssp)$, the integrated observable $\Obser^\zeit$ is the
displacement $\ssp(\zeit)$ in $d$ dimensions, and $\Delta/2d$ is
Einstein's diffusion coefficient.

In order to obtain $s(\beta)$, we construct the resolvent of \Aop , by taking
the Laplace transform of \refeq{eq-EvOpExp}:
\beq
        \int_0^{\infty} \!d\zeit\, e^{-s\zeit} \Lop^\zeit = (s-\Aop)^{-1} \, ,
        \label{eq-ResolventA}
\eeq
the trace of which peaks at the eigenvalues of \Aop. By taking the
Laplace transform of $\Lop^\zeit$ and computing its trace
by $\tr \Lop^\zeit = \int d\ssp \, \Lop^\zeit (\ssp,\ssp)$, one obtains the
classical trace formula\rf{pexp}:
\beq
\sum_{\alpha=0}^{\infty} \frac{1}{s-s_{\alpha}} = \sum_p \period{p}
\sum_{r=1}^{\infty} \frac{e^{r(\beta \Obser_p - s \period{p})}}{\oneMinJ{r}}
\ee{e-ClassicalTraceFormula}
that relates the spectrum of the {\evOper} to the spectrum of periodic
orbits. Here  $s$ is the auxiliary
variable of the Laplace transform and $s_{\alpha}$ are the eigenvalues of \Aop . The
outer sum on the right hand side runs over the `prime cycles' $p$ of the system,
\ie, the shortest \po s of period $\period{p}$. $\Obser_p$ is the value of
the observable integrated along the prime cycle and $\monodromy_p$ is the
transverse monodromy matrix, the eigenvalues $\Obser^\zeit (\ssp)$ of
which are the Floquet multipliers of $p$ with the marginal ones excluded.
In the derivation of \refeq{e-ClassicalTraceFormula}, one assumes that
the flow has a single marginal direction, namely the $\vel(\ssp)$ tangent
to the periodic orbit, and evaluates the contribution of each \po\ to the
trace integral by transforming to a local coordinate system where one of
the coordinates is parallel to the flow, while the rest are transverse.
The integral along the parallel direction contributes the factors of
$\period{p}$ in \refeq{e-ClassicalTraceFormula}. The transverse integral
over the delta function \refeq{e-EvOper} contributes the factor of
$1/\oneMinJ{r}$.

\subsection{Decomposition of the trace formula over irreducible representations}
\label{s-ContFac}

The classical trace formula \refeq{e-ClassicalTraceFormula} accounts for contributions from \po s
to long time dynamical averages. However, \rpo s of equivariant systems are almost never
periodic in the full \statesp. In order to compute the contributions of \rpo s
to the trace of the \evOper, one has to factorize
the \evOper\
into the irreducible subspaces of the symmetry group. For discrete symmetries,
this procedure is studied in \refref{CvitaEckardt}. For the quantum systems
with continuous symmetries (abelian and 3D rotations), the factorization of
the semiclassical Green's operator is carried out in \refref{Creagh93}.
Reference \cite{Cvi07} addresses the continuous factorization of the \evOper\ and its
trace; we provide a sketch of this treatment here. 

We start by stating, without
proof, that a square-integrable field $\psi (\ssp)$ over a vector space can be
factorized into its projections over the irreducible subspaces of a group
$\Group$:
\beq
    \psi (\ssp) = \sum_m \mathbb{P}_m \psi (\ssp) \, ,
\eeq
where the sum runs over the irreducible representations of $\Group$ and
the projection operator onto the $m$th irreducible subspace, for a continuous
group, is
\beq
    \mathbb{P}_m = d_m \int_\Group d \mu(\LieEl) \chi_m (\LieEl(\theta))
                            \mathbb{D}(\theta)
\,.
\ee{e-ProjectionOperator}
Here, $d_m$ is the dimension of the representation, $d \mu(g)$ is the
normalized Haar measure, $\chi_m (\LieEl)$ is the character of $m$th
irreducible representation, and $\mathbb{D}(\theta)$ is the operator that
transforms a scalar field defined on the \statesp\ as
$\mathbb{D}(\theta) \rho (\ssp) = \rho(\matrixRep(\theta)^{-1} \ssp)$.
For our specific case of a single $\SOn{2}$ symmetry,
\bea
d_m &\rightarrow& 1\, , \\
\int_G d \mu(g) &\rightarrow& \oint \frac{d \theta} {2 \pi} \, , \\
\chi_m (\LieEl(\theta)) &\rightarrow& e^{- \ii m \theta } \, .
\eea
Because the projection
operator \refeq{e-ProjectionOperator} decomposes scalar fields defined over the \statesp\
into their irreducible subspaces under action of $\Group$, it can be used to
factorize the \evOper. Thus, the kernel
of the \evOper\ transforms under the action of $\mathbb{D}(\theta)$ as
\bea
    \mathbb{D}(\theta) \Lop^\zeit (\ssp', \ssp) &=&
        \Lop^\zeit (\matrixRep(\theta)^{-1} \ssp', \ssp)\,,
    \continue
    &=& \Lop^\zeit (\ssp', \matrixRep(\theta) \ssp) \,, \continue
    &=& \delta (\ssp' - \matrixRep(\theta) f^\zeit (\ssp)) e^{\beta \Obser^\zeit(\ssp)}\, ,
    \label{e-gEvOper}
\eea
where the second step follows from the equivariance of the system under
consideration. \Rpo s contribute to $\mathbb{P}_m \Lop^\zeit = \Lop_m^\zeit$ since when its
kernel is modified as in \refeq{e-gEvOper}, the projection involves an integral
over the group parameters that is non-zero when $\theta= - \theta_{\rpprime}$, the phase shifts of the
\rpo s. By computing the trace of $\Lop_m^\zeit$, which in addition to the integral
over \statesp , now involves another integral over the group parameters, one
obtains the $m$th irreducible subspace contribution to the classical trace as
\beq
\sum_{\alpha=0}^{\infty} \frac{1}{s-s_{m, \alpha}} = \sum_p T_{\rpprime}
\sum_{r=1}^{\infty} \frac{\chi_m (\LieEl^r(\theta_{\rpprime}))
            e^{r(\beta \Obser_{\rpprime} - s T_{\rpprime})}}{\oneMinJred{r}} .
\ee{e-ReducedTraceFormula}
The reduced trace formula \refeq{e-ReducedTraceFormula} differs from the
classical trace formula \refeq{e-ClassicalTraceFormula} by the group character
term, which is evaluated at the \rpo\ phase shifts, and the reduced monodromy
matrix $\monodromyRed$, which is the $(d-N-1)\times(d-N-1)$ reduced Jacobian
for the \rpo\ evaluated on a Poincar\'e section in the \reducedsp . The eigenvalues
of $\monodromyRed$ are those of the \rpo\ Jacobian \refeq{e-rpoJacobian}
excluding the marginal ones, i.e., the ones corresponding to time evolution and evolution
along the continuous symmetry directions.

Since we are only interested in the leading eigenvalue of the \evOper , we
only consider contributions to the
trace \refeq{e-ClassicalTraceFormula} from the projections
\refeq{e-ReducedTraceFormula} of the $0$th irreducible subspace. For the $\SOn{2}$ case at hand, these can be written
explicitly as
\beq
\sum_{\alpha=0}^{\infty} \frac{1}{s-s_{0, \alpha}} = \sum_p \period{p}
\sum_{r=1}^{\infty} \frac{e^{r(\beta \Obser_p - s \period{p})}}{\oneMinJred{r}} \, .
\ee{e-tracem0}
This form differs from the classical trace formula
\refeq{e-ClassicalTraceFormula} only by the use of the reduced monodromy matrix
instead of the full monodromy matrix since
the $0$th irreducible representation of $\SOn{2}$ has character $1$. For this
reason, cycle expansions,\rf{AACI} which we cover next, are applicable
to \refeq{e-tracem0} after the replacement
$\monodromy \rightarrow \monodromyRed$.

\subsection{Cycle expansions}
\label{s-CycExp}

While the classical trace formula \refeq{e-ClassicalTraceFormula} and its
factorization for systems with continuous symmetry \refeq{e-ReducedTraceFormula} manifest
the essential duality between the spectrum of an observable and that of
the \po s and \rpo s, in practice, they are hard to work with since the
eigenvalues are located at the poles of \refeq{e-ClassicalTraceFormula} and
\refeq{e-ReducedTraceFormula}. The dynamical zeta function
\refeq{e-DynamicalZeta}, which we derive below, provides a perturbative expansion form that
enables us to order terms in decreasing importance while computing
spectra for the \twomode\ system. As stated earlier, \refeq{e-tracem0}
is equivalent to \refeq{e-ClassicalTraceFormula} via substitution
$\monodromy \rightarrow \monodromyRed$, so this derivation works for either. 

We start by defining the
`spectral determinant':
\beq
  \det (s-\Aop) = \exp \left( - \sum_p \sum_{r=1}^{\infty}
      \frac{1}{r} \frac{e^{r(\beta \Obser_p - s \period{p})}}{\oneMinJ{r}} \right)\, ,
\ee{e-SpectralDeterminant}
whose logarithmic derivative ($(d/ds) \ln \det(s - \Aop)$) gives
the classical trace formula \refeq{e-ClassicalTraceFormula}.
The spectral determinant \refeq{e-SpectralDeterminant} is easier to work
with since the spectrum of $\mathcal{A}$ is now located at the zeros of
\refeq{e-SpectralDeterminant}. The convergence of \refeq{e-SpectralDeterminant}
is, however, still not obvious. More insight is gained by approximating
$\oneMinJ{r}$ by the product of expanding Floquet multipliers and then
carrying out the sum over $r$ in \refeq{e-SpectralDeterminant}. This
approximation yields
\bea
\oneMinJ{} &=& | (1 - \ExpaEig_{e,1})(1 - \ExpaEig_{e,2})... \continue
                        &&(1 - \ExpaEig_{c,1}) (1 - \ExpaEig_{c,2}) ... | \nonumber \\
                        &\approx& \prod_e |\ExpaEig_e| \equiv |\ExpaEig_p|,
    \label{e-LambdapApprox}
\eea
where $|\ExpaEig_{e,i}| > 1$ and $|\ExpaEig_{c,i}| < 1$ are expanding and
contracting Floquet multipliers respectively. By making this approximation, the sum over $r$ in
\refeq{e-SpectralDeterminant} becomes the Taylor expansion of natural logarithm. Carrying out this sum, brings the
spectral determinant \refeq{e-SpectralDeterminant} to a product (over prime
cycles) known as the dynamical zeta function:
\beq
1 / \zeta = \prod_p (1 - t_p) \, \mbox{where}, \, t_p = \frac{1}{|\ExpaEig_p|}
            e^{\beta \Obser_p - s \period{p}} z^{n_p} .
\ee{e-DynamicalZeta}
Each `cycle weight' $t_p$ is multiplied by the `order tracking term' $z^{n_p}$,
where $n_p$ is the topological length of the $p$th prime cycle. This polynomial
ordering arises naturally in the study of discrete time systems where the
Laplace transform is replaced by $z$-transform. Here, we insert the powers of
$z$ by hand, to keep track of the ordering, and then set its value to $1$ at the
end of calculation. Doing so allows us to write the dynamical zeta function
\refeq{e-DynamicalZeta} in the `cycle expansion' form by grouping its
terms in powers of $z$. For complete binary symbolic dynamics, where every binary symbol
sequence is accessible, the cycle expansion reads
\bea
1 / \zeta &=& 1 - t_0 - t_1 - (t_{01} - t_0 t_1 )  \label{e-CycleExpansion} \\
                  && - [(t_{011} - t_{01}t_1) + (t_{001} - t_{01} t_0)] - ... \continue
                  &=& 1 - \sum_f t_f - \sum_n \hat{c}_n \label{e-CurvatureExpansion},
\eea
where we labeled each prime cycle by its binary symbol sequence. In
\refeq{e-CurvatureExpansion} we grouped the contributions to the zeta function
into two groups: `fundamental' contributions $t_f$ and `curvature' corrections $c_n$.
The curvature correction terms are denoted explicitly by parentheses in \refeq{e-CycleExpansion} and
correspond to `shadowing' combinations where combinations of
shorter cycle weights, also known as `pseudocycle' weights, are subtracted from the weights of longer
prime cycles. Since the cycle weights in \refeq{e-DynamicalZeta} already
decrease exponentially with increasing cycle period, the cycle expansion
\refeq{e-CycleExpansion} converges even faster than exponentially when the
terms corresponding to longer prime cycles are shadowed.

For complete binary symbolic dynamics, the only fundamental contributions to
the dynamical zeta function are from the cycles with topological length $1$, and all
longer cycles appear in the shadowing pseudocycle combinations.
More generally, if the symbolic dynamics is a subshift of finite type,\rf{DasBuch}
with the grammar of admissible sequences described by a finite set of pruning rules,
and the flow is uniformly hyperbolic, cycle expansions of {\Fd s}
are guaranteed to converge super-exponentially.\rf{hhrugh92}
A generic unimodal map symbolic dynamics is not a subshift of finite type.
However, we have shown in \refsect{s:numerics} that the Poincar\'e return map for
the \twomode\ system (\reffig{fig:psectandretmap}\,(d)) diverges at
$s \approx 0.98$ and approximated it as if its tip was located at the
furthest point visited by an ergodic trajectory. This brings the question of
whether we can approximate the map in \reffig{fig:psectandretmap}\,(d) in such a way
that corresponding symbolic dynamics has a finite grammar of pruning rules?
The answer is yes.

As shown in \reffig{fig:psectandretmap}\,(d), the cycles \cycle{001}
and \cycle{011} pass quite close to the tip of the cusp. Approximating the
map as if its tip is located exactly at the point where \cycle{001} cuts gives us
what we are looking for: a single grammar rule, which says that the symbol
sequence `00' is inadmissible. This can be made rigorous by the help of
kneading theory, however, the simple result is easy to see from the return map
in \reffig{fig:psectandretmap}\,(d): Cover the parts of the return map, which
are outside the borders set by the red dashed lines, the cycle \cycle{001} and
then start any point to the left of the tip and look at images. You will always
land on a point to the right of the tip, unless you start at the lower left
corner, exactly on the cycle \cycle{001}. As we will show, this `finite grammar
approximation' is reasonable since the orbits that visit outside
the borders set by \cycle{001} are very unstable, and hence, less
important for the description of invariant dynamics.

The binary grammar with only rule that forbids repeats of one of the symbols is
known as the `golden mean' shift,\rf{DasBuch} because it has a topological entropy of
$\ln \left(\left(1 + \sqrt{5}\right)/2\right)$. Binary itineraries of golden mean cycles can be easily
obtained from the complete binary symbolic dynamics by substitution
$0 \rightarrow 01$ in  the latter. Thus, we can write the dynamical zeta
function for the golden mean pruned symbolic dynamics by replacing $0$s in
\refeq{e-CycleExpansion} by $01$:
\bea
1 / \zeta &=& 1 - t_{01} - t_1 - (t_{011} - t_{01} t_1 )
              \label{e-GoldenMeanCycleExpansion}\\
                  && - [(t_{0111} - t_{011}t_1) + (t_{01011} - t_{01} t_{011} ) ] - ...
          \nonumber
\eea
Note that all the contributions longer than topological length $2$ to the
golden mean dynamical zeta function are in form of shadowing combinations. In \refsect{s-NumResults},
we will compare the convergence of the cycle averages with and without the
finite grammar approximation, but before moving on to numerical results,
we explain the remaining details of computation.

While dynamical zeta functions are useful for investigating the convergence
properties, they are not exact, and their computational cost is same as that of
exact spectral determinants. For this reason, we expand the
spectral determinant \refeq{e-SpectralDeterminant} ordered in the topological
length of cycles and pseudocycles. We start with the following form of the
spectral determinant \refeq{e-SpectralDeterminant}
\beq
    \det (s-\Aop) =   \prod_p \exp \left( - \sum_{r=1}^{n_p r < N}
                             \frac{1}{r} \frac{e^{r(\beta \Obser_p - s \period{p})}
                                          }{\oneMinJ{r}} z^{n_p r} \right) \, ,
\ee{e-SpectralDeterminantExp}
where the sum over the prime cycles in the exponential becomes a
product. We also inserted the order tracking term $z$ and truncated the sum over cycle
repeats at the expansion order $N$. For each prime cycle, we compute the sum in
\refeq{e-SpectralDeterminantExp} and expand the exponential up to order
$N$. We then multiply this expansion with the contributions from previous cycles
and drop terms with order greater than $N$. This way, after setting $z=1$,
we obtain the spectral determinant truncated to cycles and pseudo-cycles of
topological length up to $\cl{p} \leq N$,
\beq
    F_N(\beta , s) = 1 - \sum_{n=1}^{N} Q_n(\beta,s) \,. %Switched arguments of Q_n for symmetry with F_N
    \label{e-NthOrderSpectDet}
\eeq
In what follows, we shall drop the subscript, $F_N \to F$, but actual
calculations are always done for a range of finite truncation lengths
$N$. Remember that we are searching for the eigenvalues $s(\beta)$ of the
operator \Aop\ in order to compute the moments \refeq{eq-moments}.
These eigenvalues are located at the zeros of the spectral
determinant, hence as  function of $\beta$ they satisfy the implicit
equation
\beq
    F(\beta, s(\beta )) = 0 \, .
    \label{e-FNimplicit}
\eeq
By taking derivative of \refeq{e-FNimplicit} with respect to $\beta$ and
applying chain rule we obtain
\beq
    \frac{d s}{d \beta} = - \left. \frac{\partial F}{\partial \beta} \right/
                                     \frac{\partial F}{\partial s}\, .
\eeq
Higher order derivatives can be evaluated similarly.
Defining
\bea
    \langle \Obser \rangle &=& -  \partial F / \partial \beta
                               \continue %\label{e-ObserAvg} \\
        \langle \period{}\,\rangle &=&  \partial F / \partial s
                    \,,\qquad
        \langle \period{}^2 \rangle \,=\,  \partial^2 F / \partial s^2
                        \label{eq-Tavg} \\
    \langle \Obser^2 \rangle &=&
        -  \partial^2 F / \partial \beta^2
          %\label{e-Obser2Avg} \\
                    \,,\qquad
    \langle \Obser \period{}\,\rangle \,=\,
           \partial^2 F / \partial \beta \partial s
          \, ,\nnu % \label{e-ObserTAvg}
\eea
we write the \cycForm s as
\bea
    \langle \obser \rangle
            &=& \langle \Obser \rangle / \langle \period{}\,\rangle \, ,
    \label{e-Avga} \\
   \Delta
            &=&
           \frac{ 1~}{\langle \period{}\,\rangle}\left(
          \langle \Obser^2 \rangle
           \,-\, 2 \frac{d s}{d\beta}
              \langle \Obser \period{}\,\rangle
           \,+\, \left(\frac{d s}{d \beta}\right)^2
                       \langle \period{}^2 \rangle \right)
           \continue
           &=&
   \frac{ 1~}{\langle \period{}\,\rangle}
   \langle (\Obser-\period{}\,\langle  \obser \rangle)^2 \rangle
    \label{e-Avgsigma} \, ,
\eea
with everything evaluated at $\beta=0$, $s=s (0)$.

By probability conservation, we expect that for an invariant measure
$\rho_0(\ssp)$, the eigenvalue $s(0)$ is $0$. However, we did not make
this substitution in \cycForm s since, in practice, our approximations to
the spectral determinant are always based on a finite number of \po s, so
that the solution of $F_N(0, s(0)) = 0$ is small, but not exactly $0$. This
eigenvalue has a special meaning: It indicates how well the \po s cover
the strange attractor. Following this interpretation, we define $\gamma =
- s(0)$ as the `escape rate': the rate at which the dynamics escape the
region that is covered by the \po s. Specifically, for our finite grammar
approximation; the escape rate tells us how frequently the ergodic flow
visits the part of the Poincar\'e map that we cut off by applying our
finite grammar approximation.

We defined $\langle T \rangle$ in \refeq{eq-Tavg} as a shorthand for a partial
derivative, however, we can also develop an interpretation for it by looking
at the definitions of the dynamical zeta function \refeq{e-DynamicalZeta} and the
spectral determinant \refeq{e-SpectralDeterminant}. In both series, the partial
derivative with respect to $s$ turns them into a sum weighted by the cycle
periods; with this intuition, we define $\langle T \rangle$ as the `mean cycle
period'.

These remarks conclude our review of periodic orbit theory and its
extension to the equivariant dynamical systems. We are now ready to present
our numerical results and discuss their quality.

\subsection{Numerical results}
\label{s-NumResults}

We constructed the spectral determinant \refeq{e-NthOrderSpectDet} to different
orders for two observables: phase velocity $\dot{\theta}$ and the leading
Lyapunov exponent. Remember that $\Obser_p$ appearing in
\refeq{e-SpectralDeterminantExp} is the integrated observable, so in order to
obtain the moments of phase velocity and the leading Lyapunov exponent from
\refeq{e-Avga} and \refeq{e-Avgsigma}, we respectively put in
$\Obser_p = \theta_p$, the phase shift of the prime cycle $p$, and
$\Obser_p = \ln |\Lambda_{p,e}|$, the logarithm of its expanding Floquet
multiplier of $\Lambda_{p,e}$.

In \refsect{s:visual}, we explained that \SOn{2} phase shifts correspond
to the drifts in configuration space. We define the
corresponding diffusion coefficient as
\beq
    D = \frac{1}{2 d}         \lim_{\zeit \rightarrow \infty} \frac{1}{\zeit}
        \langle \theta(\zeit)^2 - \langle \theta(\zeit) \rangle^2 \rangle
%      = \frac{\Delta_{\theta}}{2}
\,,
\eeq
where $d=1$ since the configuration space is one dimensional.

\reftabs{t-DynamicalAverages}{t-DynamicalAveragesNoGrammar} respectively show
the cycle averages of the escape rate $\gamma$, mean period
$\langle T \rangle$, leading Lyapunov exponent $\Lyap$, mean phase velocity
$\langle \dot{\theta} \rangle$ and the diffusion coefficient $D$
with and without the finite grammar approximation. In the latter, we input
all the \rpo s we have found into the expansion
\refeq{e-SpectralDeterminantExp}, whereas in the former, we discarded the
cycles with symbol sequence `00'.
\begin{table}
    \caption{Cycle expansion estimates for
             the escape rate $\gamma$, average cycle period $\langle T \rangle$,
             Lyapunov exponent $\lambda$, average phase velocity
             $\langle \dot{\theta} \rangle$, and the diffusion coefficient $D$,
             using cycles up to length $N$ in the golden mean approximation
             \refeq{e-GoldenMeanCycleExpansion} of the symbolic dynamics.}
    \begin{tabular}{c|c|c|c|c|c}
     $N$ & $\gamma$ & $\langle T \rangle$ & $\lambda$ & $\langle \dot{\theta} \rangle$ & $D$ \\
    \hline
    1 & 0.249829963 & 3.6415122 & 0.10834917 & 0.0222352 & 0.000000 \\
    2 & -0.011597609 & 5.8967605 & 0.10302891 & -0.1391709 & 0.143470 \\
    3 & 0.027446312 & 4.7271381 & 0.11849761 & -0.1414933 & 0.168658 \\
    4 & -0.004455525 & 6.2386572 & 0.10631066 & -0.2141194 & 0.152201 \\
    5 & 0.000681027 & 5.8967424 & 0.11842700 & -0.2120545 & 0.164757 \\
    6 & 0.000684898 & 5.8968762 & 0.11820050 & -0.1986756 & 0.157124 \\
    7 & 0.000630426 & 5.9031596 & 0.11835159 & -0.1997353 & 0.157345 \\
    8 & 0.000714870 & 5.8918832 & 0.11827581 & -0.1982025 & 0.156001 \\
    9 & 0.000728657 & 5.8897511 & 0.11826873 & -0.1982254 & 0.156091 \\
    10 & 0.000728070 & 5.8898549 & 0.11826788 & -0.1982568 & 0.156217 \\
    11 & 0.000727891 & 5.8898903 & 0.11826778 & -0.1982561 & 0.156218 \\
    12 & 0.000727889 & 5.8898908 & 0.11826780 & -0.1982563 & 0.156220 \\
    \end{tabular}
    \label{t-DynamicalAverages}
\end{table}

\begin{table}
    \caption{Cycle expansion estimates of the escape rate $\gamma$, average
    cycle period $\langle T \rangle$, Lyapunov exponent $\lambda$, average
    phase velocity $\langle \dot{\theta} \rangle$, and the diffusion coefficient
    $D$ using all cycles found up to length $N$.}
    \label{t-DynamicalAveragesNoGrammar}
    \begin{tabular}{c|c|c|c|c|c}
     $N$ & $\gamma$ & $\langle T \rangle$ & $\lambda$ & $\langle \dot{\theta} \rangle$ & $D$ \\
    \hline
    1 & 0.249829963 & 3.6415122 & 0.10834917 & 0.0222352 & 0.000000 \\
    2 & -0.011597609 & 5.8967605 & 0.10302891 & -0.1391709 & 0.143470 \\
    3 & 0.022614694 & 4.8899587 & 0.13055574 & -0.1594782 & 0.190922 \\
    4 & -0.006065601 & 6.2482261 & 0.11086469 & -0.2191881 & 0.157668 \\
    5 & 0.000912644 & 5.7771642 & 0.11812034 & -0.2128347 & 0.168337 \\
    6 & 0.000262099 & 5.8364534 & 0.11948918 & -0.2007615 & 0.160662 \\
    7 & 0.000017707 & 5.8638210 & 0.12058951 & -0.2021046 & 0.160364 \\
    8 & 0.000113284 & 5.8511045 & 0.12028459 & -0.2006143 & 0.159233 \\
    9 & 0.000064082 & 5.8587350 & 0.12045664 & -0.2006756 & 0.158234 \\
    10 & 0.000093124 & 5.8536181 & 0.12035185 & -0.2007018 & 0.158811 \\
    11 & 0.000153085 & 5.8417694 & 0.12014700 & -0.2004520 & 0.158255 \\
    12 & 0.000135887 & 5.8455331 & 0.12019940 & -0.2005299 & 0.158465 \\
    \end{tabular}
\end{table}
In \refsect{s-CycExp}, we motivated the finite grammar approximation by
claiming that it would lead to faster convergence of dynamical averages 
due to the nearly exact shadowing combinations of the golden mean zeta function
\refeq{e-GoldenMeanCycleExpansion}. This claim is supported by the data
in \reftabs{t-DynamicalAverages}{t-DynamicalAveragesNoGrammar}. Take, for example, the
Lyapunov exponent. This converges to $7$ digits for the $12^{\mathrm{th}}$ order
expansion when using the finite grammar approximation in
\reftab{t-DynamicalAverages}, but only converges to $4$ digits at this order in
\reftab{t-DynamicalAveragesNoGrammar}. Other observables compare similarly in
terms of their convergence in both cases. Note, however, that the escape rate
in \reftab{t-DynamicalAverages} converges to $\gamma = 0.000727889$, whereas
in \reftab{t-DynamicalAveragesNoGrammar} it gets smaller and smaller with an
oscillatory behavior. This is due to the fact that in the finite grammar
approximation, we threw out the part of attractor that corresponds to the
cusp of the return map in \reffig{fig:psectandretmap}\,(d) above the point cut 
by \cycle{001}.

In order to compare with the cycle averages, we numerically estimated the
leading Lyapunov exponent of the \twomode\ system using the method of
Wolf \etal\rf{WolfSwift85} This procedure was repeated 100 times for
different initial conditions, yielding a numerical mean estimate of
$\timeAver{\Lyap} = 0.1198 \pm 0.0008$. While the finite grammar
estimate $\Lyap_{FG} = 0.1183$ is within $0.6\%$ range of this value,
the full cycle expansion agrees with the numerical estimate. This is not
surprising since in the finite grammar approximation, we discard the
most unstable cycles to obtain faster convergence, and so can expect 
a slight underestimate of the Lyapunov exponent.

\renewcommand{\zeit}{\ensuremath{\tau}}  %time variable

\section{Conclusions and discussion}
\label{s:concl}

In this tutorial, we have studied a simple dynamical 
system that exhibits chaos and is equivariant under a continuous 
symmetry transformation. We have shown that reducing this symmetry 
simplifies the qualitative dynamics to a great extent and enables 
one to find all \rpo s of the systems via standard techniques such 
as Poincar\'e sections and return maps. In addition, we have shown 
that one can extract quantitative information from the \rpo s by 
computing cycle averages.

We motivated our study of the \twomode\ system by the resemblance of its
symmetry structure to that of spatially extended systems. The
steps outlined here are, in principle, applicable to physical systems
that are described by $N$-Fourier mode truncations of PDEs such as $1D$
\KS,\rf{SCD07} $3D$ pipe flows,\rf{WiShCv15} \etc

In \refsect{s:numerics}, we showed that the dynamics of our \twomode\ model 
can be completely described by a unimodal return map of the Poincar\'e 
section that we constructed after continuous symmetry reduction. In a 
high-dimensional system, finding such an easy symbolic dynamics, 
or any symbolic dynamics at all is a challenging problem on its 
own. In \refref{lanCvit07}, the authors found that for 
the desymmetrized (confined in the odd subspace) $1$D 
spatio-temporally chaotic \KS\ system a bimodal return map could 
be obtained after reducing the discrete symmetry of the 
problem. However, we do not know any study that has been able to 
simplify turbulent fluid flow to such an extent.

In \refsect{s:DynAvers}, we showed that symbolic dynamics and their
associated grammar rules greatly affect the convergence of \cycForm s.
In general, finding a finite symbolic description of a flow is
rarely as easy as it is in our model system.
There exist other methods of ordering cycle
expansion terms, for example, ordering pseudo-cycles by their stability and discarding terms
that are above a threshold.\rf{DM97} In this case, one expects the remaining terms to form
shadowing combinations and converge exponentially.
Whichever method of term ordering is employed, the cycle expansions are only as good
as the least unstable cycle that one fails to find. Symbolic dynamics solves both 
problems at once since it puts the cycles in order by topological length so that
one cannot miss any accessible cycle and shadowing combinations occur naturally. 
The question one might then ask is: When there is no symbolic dynamics, how can you make 
sure that you have found all the periodic orbits of a flow up to some cycle period?

In searching for cycles in high-dimensional flows, one usually looks at
the near recurrences of the ergodic flow and then runs Newton searches
starting near these recurrences to find if they are influenced by a nearby exactly
recurrent solution. Such an approach does not answer the question we just asked with full
confidence, although one might argue that the dynamically important cycles
influence the ergodic flow, leading to recurrences, and thus, cycles found this way are
those that are relevant for computing averages.

To sum up, we have shown that periodic orbit theory successfully 
extends to systems with continuous symmetries. When dealing with 
high dimensional systems, one still needs to think about some of the 
remaining challenges outlined above. Once these are overcome, it 
should become possible to extract quantitative information about 
turbulence by using exact unstable solutions of the Navier-Stokes 
equations.

\begin{acknowledgments}
We are grateful to Evangelos Siminos for his contributions to this project
and Mohammad M.~Farazmand for a critical reading of the manuscript.
We acknowledge stimulating discussion with
Xiong Ding,
Ruslan L.~Davidchack,
Ashley P.~Willis,
Al Shapere
and
Francesco Fedele.
We are indebted to the 2012 ChaosBook.org class, in particular to
Keith M.~Carroll,
Sarah Flynn,
Bryce Robbins,
and
Lei Zhang,
for the initial fearless fishing expeditions into the enormous sea of
parameter values of the \twomode\ model.
P.~C.\ thanks the family of late G.~Robinson,~Jr.
and
NSF~DMS-1211827 for support. D.~B.\ thanks M.~F.\ Schatz for support during
the early stages of this work under NSF~CBET-0853691.
\end{acknowledgments}

\appendix

\section{Multiple shooting method for finding \rpo s}
\label{s:newton}

Let us assume that we have a set of good guesses
for a set of \statesp\ points,
flight times and
$1D$ symmetry group parameter increments
$\{\ssp_i^{(0)}\,,\,\zeit_i^{(0)}\,,\,\gSpace_i^{(0)}\}$
such that the points
$\{\ssp_i^{(0)}\}$
lie close to the \rpo\ $p$ such that
\beq
        \ssp_{i+1}^{(0)}
\approx
    \matrixRep(- \gSpace_i^{(0)}) \flow{{\zeit_i^{(0)}}}{\ssp_i^{(0)}}
\quad
    \mbox{cyclic in $i = 1, ..., n$}
\,.
\eeq
Here, the period and the shift of the \rpo\ $p$ are
$\period{p} \approx \sum \zeit_i\,,$ and
$\gSpace_p \approx \sum \gSpace_i$. The Lagrangian description of the 
flow is then $\ssp(\zeit) = \flow{\zeit}{\ssp(0)}$.
We want to determine corrections
$(\Delta\ssp_i\,,\,\Delta\zeit_i\,,\,\Delta\gSpace_i)$ so that
\bea
        \ssp_{i+1} + \Delta \ssp_{i+1} &=& \matrixRep(- \gSpace_i - \Delta \gSpace_i)
                \flow{\zeit_i + \Delta \zeit_i}{\ssp_i + \Delta \ssp_i} \continue
                &&  \mbox{cyclic in } i = 1, ..., n
\,.
\eea
To linear order in
\bea
&& (\Delta\ssp_i^{(m+1)}\,,\, \Delta\zeit_i^{(m+1)}\,,\,\Delta\gSpace_i^{(m+1)}) \\
&&=
(\ssp_i^{(m+1)}-\ssp_i^{(m)}\,,\,
 \zeit_i^{(m+1)}-\zeit_i^{(m)}\,,\,
 \gSpace_i^{(m+1)}-\gSpace_i^{(m)}) \nonumber
\eea
the improved Newton guess
$
(\ssp_i^{(m+1)}\,,\,\zeit_i^{(m+1)}\,,\,\gSpace_i^{(m+1)} )
$
is obtained by minimizing the effect of perturbations along
the spatial, time, and phase directions,
\bea
        && \ssp_{i+1}^{'} - \matrixRep_{i+1} \flow{{\zeit_i}}{\ssp_i} \continue
        && = \matrixRep_{i+1}\left(
                                                           \jMps_{i+1} \Delta \ssp_i
                                                         + \vel_{i+1} \Delta \zeit_i
                                                         - \groupTan_{i+1} \Delta \gSpace_i
                                         \right) \,,
\label{PCnetwonStep}
\eea
where, for brevity,
$\ssp_{i}^{(m+1)} = \ssp_{i}^{(m)} + \Delta \ssp_{i}^{(m)}
   = \ssp_{i}^{'}$,
$\ssp_{i}^{(m)} = \ssp_{i}$,
$\matrixRep(- \gSpace_i) = \matrixRep_{i+1}$,
$\vel (\ssp_{i}(\zeit_{i})) = \vel_{i+1}$,
$\jMps^{\zeit_i}(\ssp_i) = \jMps_{i+1}$,
$\groupTan(\ssp_{i}(\zeit_{i})) = \Lg \ssp_{i}(\zeit_{i}) = \groupTan_{i+1}$,
\etc %\etc already has period in it.
For sufficiently good initial guesses,
the improved values converge under Newton iterations to
the exact values
$(\Delta\ssp_i\,,\,\Delta\zeit_i\,,\,\Delta\gSpace_i)$
=$\,(\Delta\ssp_i^{(\infty)}\,,\,\Delta\zeit_i^{(\infty)}\,,\,\Delta\gSpace_i^{(\infty)})$
at a super-exponential rate.

In order to deal with the marginal multipliers along the time and group
orbit directions, one needs to apply a pair of constraints, which
eliminate variations along the marginal directions on the \rpo's $2D$
torus. These can be formulated as a local Poincar\'e section orthogonal 
to the flow and a local slice orthogonal to the group orbit at each point 
along the orbit,
\beq
   \braket{\vel(\ssp_i )}{\Delta \ssp_i} = 0
\,,\qquad
   \braket{\groupTan(\ssp_i )}{\Delta \ssp_i} = 0
\,.
\ee{RPOConstrsLocal}
We can rewrite everything as one matrix equation:
\beq \label{eq:multishootmatrix}
        A \Delta = E \, ,
\eeq
where
\begin{widetext}
\bea 
        A &=& \left(
        \begin{array}{ccccccccccc}      
          \matrixRep_{2} \jMps_{2} &
          \matrixRep_{2} \vel_2 &
          - \Lg \matrixRep_{2} \flow{\zeit_1}{\ssp_1} &
          - \matId & 0 & 0 & 0 & \cdots & 0 & 0 & 0 \\
          \vel(\ssp_1) & 0 & 0 & 0 & 0 & 0 & 0 & \cdots & 0 & 0 & 0 \\
          \groupTan(\ssp_1) & 0 & 0 & 0 & 0 & 0 & 0 & \cdots & 0 & 0 & 0 \\
          0 & 0 & 0 &
          \matrixRep_{3} \jMps_{3} &
          \matrixRep_{3} \vel_3 &
          - \Lg \matrixRep_{3} \flow{\zeit_2}{\ssp_2}   &
          - \matId & \cdots & 0 & 0 & 0\\
          0 & 0 & 0 & \vel(\ssp_2) & 0 & 0 & 0 & \cdots & 0 & 0 & 0 \\
          0 & 0 & 0 & \groupTan(\ssp_2) & 0 & 0 & 0 & \cdots & 0 & 0 & 0 \\
          \vdots & \vdots & \vdots & \vdots & \vdots & \vdots & \vdots & \ddots & \vdots & \vdots & \vdots \\
          - \matId & 0 & 0 & 0 & 0 & 0 & 0 & \cdots &
          \matrixRep_{1} \jMps_{1} &
          \matrixRep_{1} \vel_1 &
          - \Lg \matrixRep_{1} \flow{\zeit_1}{\ssp_1} \\
          0 & 0 & 0 & 0 & 0 & 0 & 0 & \cdots & \vel(\ssp_n) & 0 & 0 \\
          0 & 0 & 0 & 0 & 0 & 0 & 0 & \cdots & \groupTan(\ssp_n) & 0 & 0
        \end{array} \right) \, , \label{eq:AforNewton} \\
        \Delta &=&
         (
          \Delta \ssp_1, \,
          \Delta \zeit_1, \,
          \Delta \gSpace_1, \,
          \Delta \ssp_2, \,
          \Delta \zeit_2, \,
          \Delta \gSpace_2, \,
          \ldots , \,
          \Delta \ssp_n, \,
          \Delta \zeit_n, \,
          \Delta \gSpace_n
         )^T \, ,
         \\ 
        E &=&
         (
          \ssp_{2} - \matrixRep_2 \flow{\zeit_1}{\ssp_1} , \,
           0    , \,
           0    , \,
          \ssp_{3} - \matrixRep_3 \flow{\zeit_2}{\ssp_2} , \,
          0     , \,
          0     , \,
          \ldots , \,
          \ssp_{1} - \matrixRep_1 \flow{\zeit_n}{\ssp_n} , \,
          0     , \,
          0     
          )^T \, . \label{eq:DeltaandE}  
\eea
\end{widetext}

We then solve \refeq{eq:multishootmatrix} for $\Delta$ and update our initial
guess by adding the vector of the computed $\Delta$ values to it and iterate.

\section{Periodic Schur decomposition}
\label{s:schur}

Here, we briefly summarize the periodic eigen decomposition\rf{DingCvit14}
needed for the evaluation of Floquet multipliers for \twomode\ \po s. Due to
the non-hyperbolicity of the return map of
\reffig{fig:psectandretmap}\,(d), Floquet multipliers can easily differ
by 100s of orders of magnitude even in a model as simple as the \twomode\
system.

We obtain the Jacobian of the \rpo\ as a multiplication of short-time
Jacobians from the multiple shooting computation of \refAppe{s:newton}, so that
\bea
    \jMpsRed &=& \matrixRep_n \jMps_n  \matrixRep_{n-1} \jMps_{n-1} \, ... \, \matrixRep_1 \jMps_1  \continue
                 &=& \hat{\jMps}_n \hat{\jMps}_{n-1} \, ... \, \hat{\jMps}_1 \label{e-JacobianProduct} \\
                 && \mbox{where,}\, \hat{\jMps_i} = \matrixRep_i \jMps_i \in
                    \mathbb{R}^{4 \times 4}, i = 1,2,...,n \, . \nonumber
\eea
This Jacobian is the same as the definition in \refeq{e-rpoJacobian}
since $J_i$ and $\matrixRep_i$ commute with each other and are multiplicative 
in time and phase, respectively. In order to determine the eigenvalues of $\hat{\jMps}$, we
bring each term appearing in the product \refeq{e-JacobianProduct} into periodic,
real Schur form as follows:
\beq
    \jMpsRed_i = Q_i R_i Q_{i-1}^T \, ,
\eeq
where $Q_i$ are orthogonal matrices that satisfy the cyclic property: $Q_0 = Q_n$.
After this similarity transformation, we can define $R = R_k R_{k-1} ... R_1$ and
re-write the Jacobian as:
\beq
    \jMpsRed = Q_n R Q_n^T \, .
\eeq
The matrix $R$ is, in general, block-diagonal with $1 \times 1$ blocks for real
eigenvalues and $2 \times 2$ blocks for the complex pairs. It also has the same
eigenvalues as $\hat{\jMps}$. In our case, it is diagonal since all Floquet multipliers
are real for \rpo s of the \twomode\ system. For each \rpo , we have two marginal Floquet
multipliers corresponding to the time evolution direction and the continuous symmetry direction,
as well as one expanding and one contracting eigenvalue.

\bibliography{../../bibtex/siminos}

\begin{thebibliography}{51}%
\makeatletter
\providecommand \@ifxundefined [1]{%
 \@ifx{#1\undefined}
}%
\providecommand \@ifnum [1]{%
 \ifnum #1\expandafter \@firstoftwo
 \else \expandafter \@secondoftwo
 \fi
}%
\providecommand \@ifx [1]{%
 \ifx #1\expandafter \@firstoftwo
 \else \expandafter \@secondoftwo
 \fi
}%
\providecommand \natexlab [1]{#1}%
\providecommand \enquote  [1]{``#1''}%
\providecommand \bibnamefont  [1]{#1}%
\providecommand \bibfnamefont [1]{#1}%
\providecommand \citenamefont [1]{#1}%
\providecommand \href@noop [0]{\@secondoftwo}%
\providecommand \href [0]{\begingroup \@sanitize@url \@href}%
\providecommand \@href[1]{\@@startlink{#1}\@@href}%
\providecommand \@@href[1]{\endgroup#1\@@endlink}%
\providecommand \@sanitize@url [0]{\catcode `\\12\catcode `\$12\catcode
  `\&12\catcode `\#12\catcode `\^12\catcode `\_12\catcode `\%12\relax}%
\providecommand \@@startlink[1]{}%
\providecommand \@@endlink[0]{}%
\providecommand \url  [0]{\begingroup\@sanitize@url \@url }%
\providecommand \@url [1]{\endgroup\@href {#1}{\urlprefix }}%
\providecommand \urlprefix  [0]{URL }%
\providecommand \Eprint [0]{\href }%
\providecommand \doibase [0]{http://dx.doi.org/}%
\providecommand \selectlanguage [0]{\@gobble}%
\providecommand \bibinfo  [0]{\@secondoftwo}%
\providecommand \bibfield  [0]{\@secondoftwo}%
\providecommand \translation [1]{[#1]}%
\providecommand \BibitemOpen [0]{}%
\providecommand \bibitemStop [0]{}%
\providecommand \bibitemNoStop [0]{.\EOS\space}%
\providecommand \EOS [0]{\spacefactor3000\relax}%
\providecommand \BibitemShut  [1]{\csname bibitem#1\endcsname}%
\let\auto@bib@innerbib\@empty
%</preamble>
\bibitem [{\citenamefont {Hof}\ \emph {et~al.}(2004)\citenamefont {Hof},
  \citenamefont {van Doorne}, \citenamefont {Westerweel}, \citenamefont
  {Nieuwstadt}, \citenamefont {Faisst}, \citenamefont {Eckhardt}, \citenamefont
  {Wedin}, \citenamefont {Kerswell},\ and\ \citenamefont
  {Waleffe}}]{science04}%
  \BibitemOpen
  \bibfield  {author} {\bibinfo {author} {\bibfnamefont {B.}~\bibnamefont
  {Hof}}, \bibinfo {author} {\bibfnamefont {C.~W.~H.}\ \bibnamefont {van
  Doorne}}, \bibinfo {author} {\bibfnamefont {J.}~\bibnamefont {Westerweel}},
  \bibinfo {author} {\bibfnamefont {F.~T.~M.}\ \bibnamefont {Nieuwstadt}},
  \bibinfo {author} {\bibfnamefont {H.}~\bibnamefont {Faisst}}, \bibinfo
  {author} {\bibfnamefont {B.}~\bibnamefont {Eckhardt}}, \bibinfo {author}
  {\bibfnamefont {H.}~\bibnamefont {Wedin}}, \bibinfo {author} {\bibfnamefont
  {R.~R.}\ \bibnamefont {Kerswell}}, \ and\ \bibinfo {author} {\bibfnamefont
  {F.}~\bibnamefont {Waleffe}},\ }\bibfield  {title} {\enquote {\bibinfo
  {title} {Experimental observation of nonlinear traveling waves in turbulent
  pipe flow},}\ }\href@noop {} {\bibfield  {journal} {\bibinfo  {journal}
  {Science}\ }\textbf {\bibinfo {volume} {305}},\ \bibinfo {pages} {1594--1598}
  (\bibinfo {year} {2004})}\BibitemShut {NoStop}%
\bibitem [{\citenamefont {Poincar\'e}(1896)}]{Poinc1896}%
  \BibitemOpen
  \bibfield  {author} {\bibinfo {author} {\bibfnamefont {H.}~\bibnamefont
  {Poincar\'e}},\ }\bibfield  {title} {\enquote {\bibinfo {title} {Sur les
  solutions p\'eriodiques et le principe de moindre action},}\ }\href@noop {}
  {\bibfield  {journal} {\bibinfo  {journal} {C. R. Acad. Sci. Paris}\ }\textbf
  {\bibinfo {volume} {123}},\ \bibinfo {pages} {915--918} (\bibinfo {year}
  {1896})}\BibitemShut {NoStop}%
\bibitem [{\citenamefont {Smale}(1970)}]{Smale70I}%
  \BibitemOpen
  \bibfield  {author} {\bibinfo {author} {\bibfnamefont {S.}~\bibnamefont
  {Smale}},\ }\bibfield  {title} {\enquote {\bibinfo {title} {Topology and
  mechanics, {I}.}}\ }\href@noop {} {\bibfield  {journal} {\bibinfo  {journal}
  {Invent. Math.}\ }\textbf {\bibinfo {volume} {10}},\ \bibinfo {pages}
  {305--331} (\bibinfo {year} {1970})}\BibitemShut {NoStop}%
\bibitem [{\citenamefont {Field}(1970)}]{Field70}%
  \BibitemOpen
  \bibfield  {author} {\bibinfo {author} {\bibfnamefont {M.}~\bibnamefont
  {Field}},\ }\bibfield  {title} {\enquote {\bibinfo {title} {Equivariant
  dynamical systems},}\ }\href@noop {} {\bibfield  {journal} {\bibinfo
  {journal} {Bull. Amer. Math. Soc.}\ }\textbf {\bibinfo {volume} {76}},\
  \bibinfo {pages} {1314--1318} (\bibinfo {year} {1970})}\BibitemShut {NoStop}%
\bibitem [{\citenamefont {Ruelle}(1973)}]{ruell73}%
  \BibitemOpen
  \bibfield  {author} {\bibinfo {author} {\bibfnamefont {D.}~\bibnamefont
  {Ruelle}},\ }\bibfield  {title} {\enquote {\bibinfo {title} {Bifurcations in
  presence of a symmetry group},}\ }\href@noop {} {\bibfield  {journal}
  {\bibinfo  {journal} {Arch. Rational Mech. Anal.}\ }\textbf {\bibinfo
  {volume} {51}},\ \bibinfo {pages} {136--152} (\bibinfo {year}
  {1973})}\BibitemShut {NoStop}%
\bibitem [{\citenamefont {Golubitsky}\ and\ \citenamefont
  {Stewart}(2002)}]{golubitsky2002sp}%
  \BibitemOpen
  \bibfield  {author} {\bibinfo {author} {\bibfnamefont {M.}~\bibnamefont
  {Golubitsky}}\ and\ \bibinfo {author} {\bibfnamefont {I.}~\bibnamefont
  {Stewart}},\ }\href@noop {} {\emph {\bibinfo {title} {The Symmetry
  Perspective}}}\ (\bibinfo  {publisher} {Birkh{\"a}user},\ \bibinfo {address}
  {Boston},\ \bibinfo {year} {2002})\BibitemShut {NoStop}%
\bibitem [{\citenamefont {Field}(2007)}]{Field07}%
  \BibitemOpen
  \bibfield  {author} {\bibinfo {author} {\bibfnamefont {M.~J.}\ \bibnamefont
  {Field}},\ }\href@noop {} {\emph {\bibinfo {title} {Dynamics and Symmetry}}}\
  (\bibinfo  {publisher} {Imperial College Press},\ \bibinfo {address}
  {London},\ \bibinfo {year} {2007})\BibitemShut {NoStop}%
\bibitem [{\citenamefont {Cvitanovi\'{c}}\ \emph {et~al.}(2015)\citenamefont
  {Cvitanovi\'{c}}, \citenamefont {Artuso}, \citenamefont {Mainieri},
  \citenamefont {Tanner},\ and\ \citenamefont {Vattay}}]{DasBuch}%
  \BibitemOpen
  \bibfield  {author} {\bibinfo {author} {\bibfnamefont {P.}~\bibnamefont
  {Cvitanovi\'{c}}}, \bibinfo {author} {\bibfnamefont {R.}~\bibnamefont
  {Artuso}}, \bibinfo {author} {\bibfnamefont {R.}~\bibnamefont {Mainieri}},
  \bibinfo {author} {\bibfnamefont {G.}~\bibnamefont {Tanner}}, \ and\ \bibinfo
  {author} {\bibfnamefont {G.}~\bibnamefont {Vattay}},\ }\href@noop {} {\emph
  {\bibinfo {title} {Chaos: Classical and Quantum}}}\ (\bibinfo  {publisher}
  {Niels Bohr Inst.},\ \bibinfo {address} {Copenhagen},\ \bibinfo {year}
  {2015})\ \bibinfo {note} {{\wwwcb{}}}\BibitemShut {NoStop}%
\bibitem [{\citenamefont {Gilmore}\ and\ \citenamefont
  {Letellier}(2007)}]{GL-Gil07b}%
  \BibitemOpen
  \bibfield  {author} {\bibinfo {author} {\bibfnamefont {R.}~\bibnamefont
  {Gilmore}}\ and\ \bibinfo {author} {\bibfnamefont {C.}~\bibnamefont
  {Letellier}},\ }\href@noop {} {\emph {\bibinfo {title} {The Symmetry of
  Chaos}}}\ (\bibinfo  {publisher} {Oxford Univ. Press},\ \bibinfo {address}
  {Oxford},\ \bibinfo {year} {2007})\BibitemShut {NoStop}%
\bibitem [{\citenamefont {Gatermann}(2000)}]{gatermannHab}%
  \BibitemOpen
  \bibfield  {author} {\bibinfo {author} {\bibfnamefont {K.}~\bibnamefont
  {Gatermann}},\ }\href@noop {} {\emph {\bibinfo {title} {Computer Algebra
  Methods for Equivariant Dynamical Systems}}}\ (\bibinfo  {publisher}
  {Springer},\ \bibinfo {address} {New York},\ \bibinfo {year}
  {2000})\BibitemShut {NoStop}%
\bibitem [{\citenamefont {Koenig}(1997)}]{Koenig97}%
  \BibitemOpen
  \bibfield  {author} {\bibinfo {author} {\bibfnamefont {M.}~\bibnamefont
  {Koenig}},\ }\bibfield  {title} {\enquote {\bibinfo {title} {Linearization of
  vector fields on the orbit space of the action of a compact {Lie} group},}\
  }\href {\doibase 10.1017/S0305004196001314} {\bibfield  {journal} {\bibinfo
  {journal} {Math. Proc. Cambridge Philos. Soc.}\ }\textbf {\bibinfo {volume}
  {121}},\ \bibinfo {pages} {401--424} (\bibinfo {year} {1997})}\BibitemShut
  {NoStop}%
\bibitem [{\citenamefont {Rowley}\ and\ \citenamefont
  {Marsden}(2000)}]{rowley_reconstruction_2000}%
  \BibitemOpen
  \bibfield  {author} {\bibinfo {author} {\bibfnamefont {C.~W.}\ \bibnamefont
  {Rowley}}\ and\ \bibinfo {author} {\bibfnamefont {J.~E.}\ \bibnamefont
  {Marsden}},\ }\bibfield  {title} {\enquote {\bibinfo {title} {Reconstruction
  equations and the {Karhunen-Lo\'eve} expansion for systems with symmetry},}\
  }\href@noop {} {\bibfield  {journal} {\bibinfo  {journal} {Physica D}\
  }\textbf {\bibinfo {volume} {142}},\ \bibinfo {pages} {1--19} (\bibinfo
  {year} {2000})}\BibitemShut {NoStop}%
\bibitem [{\citenamefont {Beyn}\ and\ \citenamefont
  {Th\"ummler}(2004)}]{BeTh04}%
  \BibitemOpen
  \bibfield  {author} {\bibinfo {author} {\bibfnamefont {W.-J.}\ \bibnamefont
  {Beyn}}\ and\ \bibinfo {author} {\bibfnamefont {V.}~\bibnamefont
  {Th\"ummler}},\ }\bibfield  {title} {\enquote {\bibinfo {title} {Freezing
  solutions of equivariant evolution equations},}\ }\href@noop {} {\bibfield
  {journal} {\bibinfo  {journal} {SIAM J. Appl. Dyn. Syst.}\ }\textbf {\bibinfo
  {volume} {3}},\ \bibinfo {pages} {85--116} (\bibinfo {year}
  {2004})}\BibitemShut {NoStop}%
\bibitem [{\citenamefont {Siminos}\ and\ \citenamefont
  {Cvitanovi{\'c}}(2011)}]{SiCvi10}%
  \BibitemOpen
  \bibfield  {author} {\bibinfo {author} {\bibfnamefont {E.}~\bibnamefont
  {Siminos}}\ and\ \bibinfo {author} {\bibfnamefont {P.}~\bibnamefont
  {Cvitanovi{\'c}}},\ }\bibfield  {title} {\enquote {\bibinfo {title}
  {Continuous symmetry reduction and return maps for high-dimensional flows},}\
  }\href {\doibase 10.1016/j.physd.2010.07.01} {\bibfield  {journal} {\bibinfo
  {journal} {Physica D}\ }\textbf {\bibinfo {volume} {240}},\ \bibinfo {pages}
  {187--198} (\bibinfo {year} {2011})}\BibitemShut {NoStop}%
\bibitem [{\citenamefont {Froehlich}\ and\ \citenamefont
  {Cvitanovi{\'c}}(2012)}]{FrCv11}%
  \BibitemOpen
  \bibfield  {author} {\bibinfo {author} {\bibfnamefont {S.}~\bibnamefont
  {Froehlich}}\ and\ \bibinfo {author} {\bibfnamefont {P.}~\bibnamefont
  {Cvitanovi{\'c}}},\ }\bibfield  {title} {\enquote {\bibinfo {title}
  {Reduction of continuous symmetries of chaotic flows by the method of
  slices},}\ }\href {\doibase 10.1016/j.cnsns.2011.07.007} {\bibfield
  {journal} {\bibinfo  {journal} {Commun. Nonlinear Sci. Numer. Simul.}\
  }\textbf {\bibinfo {volume} {17}},\ \bibinfo {pages} {2074--2084} (\bibinfo
  {year} {2012})},\ \bibinfo {note} {\arXiv{1101.3037}}\BibitemShut {NoStop}%
\bibitem [{\citenamefont {Cvitanovi\'c}\ \emph {et~al.}(2012)\citenamefont
  {Cvitanovi\'c}, \citenamefont {Borrero-Echeverry}, \citenamefont {Carroll},
  \citenamefont {Robbins},\ and\ \citenamefont {Siminos}}]{atlas12}%
  \BibitemOpen
  \bibfield  {author} {\bibinfo {author} {\bibfnamefont {P.}~\bibnamefont
  {Cvitanovi\'c}}, \bibinfo {author} {\bibfnamefont {D.}~\bibnamefont
  {Borrero-Echeverry}}, \bibinfo {author} {\bibfnamefont {K.}~\bibnamefont
  {Carroll}}, \bibinfo {author} {\bibfnamefont {B.}~\bibnamefont {Robbins}}, \
  and\ \bibinfo {author} {\bibfnamefont {E.}~\bibnamefont {Siminos}},\
  }\bibfield  {title} {\enquote {\bibinfo {title} {Cartography of
  high-dimensional flows: {A} visual guide to sections and slices},}\ }\href
  {\doibase 10.1063/1.4758309} {\bibfield  {journal} {\bibinfo  {journal}
  {Chaos}\ }\textbf {\bibinfo {volume} {22}},\ \bibinfo {pages} {047506}
  (\bibinfo {year} {2012})}\BibitemShut {NoStop}%
\bibitem [{\citenamefont {Willis}, \citenamefont {Cvitanovi{\'c}},\ and\
  \citenamefont {Avila}(2013)}]{ACHKW11}%
  \BibitemOpen
  \bibfield  {author} {\bibinfo {author} {\bibfnamefont {A.~P.}\ \bibnamefont
  {Willis}}, \bibinfo {author} {\bibfnamefont {P.}~\bibnamefont
  {Cvitanovi{\'c}}}, \ and\ \bibinfo {author} {\bibfnamefont {M.}~\bibnamefont
  {Avila}},\ }\bibfield  {title} {\enquote {\bibinfo {title} {Revealing the
  state space of turbulent pipe flow by symmetry reduction},}\ }\href {\doibase
  10.1017/jfm.2013.75} {\bibfield  {journal} {\bibinfo  {journal} {J. Fluid
  Mech.}\ }\textbf {\bibinfo {volume} {721}},\ \bibinfo {pages} {514--540}
  (\bibinfo {year} {2013})},\ \bibinfo {note} {\arXiv{1203.3701}}\BibitemShut
  {NoStop}%
\bibitem [{\citenamefont {Budanur}\ \emph {et~al.}(2015)\citenamefont
  {Budanur}, \citenamefont {Cvitanovi\'c}, \citenamefont {Davidchack},\ and\
  \citenamefont {Siminos}}]{BudCvi14}%
  \BibitemOpen
  \bibfield  {author} {\bibinfo {author} {\bibfnamefont {N.~B.}\ \bibnamefont
  {Budanur}}, \bibinfo {author} {\bibfnamefont {P.}~\bibnamefont
  {Cvitanovi\'c}}, \bibinfo {author} {\bibfnamefont {R.~L.}\ \bibnamefont
  {Davidchack}}, \ and\ \bibinfo {author} {\bibfnamefont {E.}~\bibnamefont
  {Siminos}},\ }\bibfield  {title} {\enquote {\bibinfo {title} {Reduction of
  the {SO(2)} symmetry for spatially extended dynamical systems},}\ }\href
  {\doibase 10.1103/PhysRevLett.114.084102} {\bibfield  {journal} {\bibinfo
  {journal} {Phys. Rev. Lett.}\ }\textbf {\bibinfo {volume} {114}},\ \bibinfo
  {pages} {084102} (\bibinfo {year} {2015})},\ \bibinfo {note}
  {\arXiv{1405.1096}}\BibitemShut {NoStop}%
\bibitem [{\citenamefont {Willis}, \citenamefont {Short},\ and\ \citenamefont
  {Cvitanovi{\'c}}(2015)}]{WiShCv15}%
  \BibitemOpen
  \bibfield  {author} {\bibinfo {author} {\bibfnamefont {A.~P.}\ \bibnamefont
  {Willis}}, \bibinfo {author} {\bibfnamefont {K.~Y.}\ \bibnamefont {Short}}, \
  and\ \bibinfo {author} {\bibfnamefont {P.}~\bibnamefont {Cvitanovi{\'c}}},\
  }\href@noop {} {\enquote {\bibinfo {title} {Relative periodic orbits form the
  backbone of turbulent pipe flow},}\ } (\bibinfo {year} {2015}),\ \bibinfo
  {note} {\arXiv{1504.05825}}\BibitemShut {NoStop}%
\bibitem [{\citenamefont {Cartan}(1935)}]{CartanMF}%
  \BibitemOpen
  \bibfield  {author} {\bibinfo {author} {\bibfnamefont {E.}~\bibnamefont
  {Cartan}},\ }\href@noop {} {\emph {\bibinfo {title} {La m\'ethode du rep\`ere
  mobile, la th\'eorie des groupes continus, et les espaces
  g\'en\'eralis\'es}}},\ \bibinfo {series} {{Expos\'es} de {G\'eom\'etrie}},
  Vol.~\bibinfo {volume} {5}\ (\bibinfo  {publisher} {Hermann},\ \bibinfo
  {address} {Paris},\ \bibinfo {year} {1935})\BibitemShut {NoStop}%
\bibitem [{\citenamefont {Field}(1980)}]{Field80}%
  \BibitemOpen
  \bibfield  {author} {\bibinfo {author} {\bibfnamefont {M.~J.}\ \bibnamefont
  {Field}},\ }\bibfield  {title} {\enquote {\bibinfo {title} {Equivariant
  dynamical systems},}\ }\href@noop {} {\bibfield  {journal} {\bibinfo
  {journal} {Trans. Amer. Math. Soc.}\ }\textbf {\bibinfo {volume} {259}},\
  \bibinfo {pages} {185--205} (\bibinfo {year} {1980})}\BibitemShut {NoStop}%
\bibitem [{\citenamefont {Krupa}(1990)}]{Krupa90}%
  \BibitemOpen
  \bibfield  {author} {\bibinfo {author} {\bibfnamefont {M.}~\bibnamefont
  {Krupa}},\ }\bibfield  {title} {\enquote {\bibinfo {title} {Bifurcations of
  relative equilibria},}\ }\href@noop {} {\bibfield  {journal} {\bibinfo
  {journal} {{SIAM} J. Math. Anal.}\ }\textbf {\bibinfo {volume} {21}},\
  \bibinfo {pages} {1453--1486} (\bibinfo {year} {1990})}\BibitemShut {NoStop}%
\bibitem [{\citenamefont {Ashwin}\ and\ \citenamefont
  {Melbourne}(1997)}]{AshMe97}%
  \BibitemOpen
  \bibfield  {author} {\bibinfo {author} {\bibfnamefont {P.}~\bibnamefont
  {Ashwin}}\ and\ \bibinfo {author} {\bibfnamefont {I.}~\bibnamefont
  {Melbourne}},\ }\bibfield  {title} {\enquote {\bibinfo {title} {Noncompact
  drift for relative equilibria and relative periodic orbits},}\ }\href@noop {}
  {\bibfield  {journal} {\bibinfo  {journal} {Nonlinearity}\ }\textbf {\bibinfo
  {volume} {10}},\ \bibinfo {pages} {595} (\bibinfo {year} {1997})}\BibitemShut
  {NoStop}%
\bibitem [{\citenamefont {Fels}\ and\ \citenamefont
  {Olver}(1998)}]{FelsOlver98}%
  \BibitemOpen
  \bibfield  {author} {\bibinfo {author} {\bibfnamefont {M.}~\bibnamefont
  {Fels}}\ and\ \bibinfo {author} {\bibfnamefont {P.~J.}\ \bibnamefont
  {Olver}},\ }\bibfield  {title} {\enquote {\bibinfo {title} {Moving coframes:
  {I. A} practical algorithm},}\ }\href@noop {} {\bibfield  {journal} {\bibinfo
   {journal} {Acta Appl. Math.}\ }\textbf {\bibinfo {volume} {51}},\ \bibinfo
  {pages} {161--213} (\bibinfo {year} {1998})}\BibitemShut {NoStop}%
\bibitem [{\citenamefont {Fels}\ and\ \citenamefont
  {Olver}(1999)}]{FelsOlver99}%
  \BibitemOpen
  \bibfield  {author} {\bibinfo {author} {\bibfnamefont {M.}~\bibnamefont
  {Fels}}\ and\ \bibinfo {author} {\bibfnamefont {P.~J.}\ \bibnamefont
  {Olver}},\ }\bibfield  {title} {\enquote {\bibinfo {title} {Moving coframes:
  {II. Regularization} and theoretical foundations},}\ }\href@noop {}
  {\bibfield  {journal} {\bibinfo  {journal} {Acta Appl. Math.}\ }\textbf
  {\bibinfo {volume} {55}},\ \bibinfo {pages} {127--208} (\bibinfo {year}
  {1999})}\BibitemShut {NoStop}%
\bibitem [{\citenamefont {Haller}\ and\ \citenamefont
  {Mezi\'c}(1998)}]{HaMe98}%
  \BibitemOpen
  \bibfield  {author} {\bibinfo {author} {\bibfnamefont {G.}~\bibnamefont
  {Haller}}\ and\ \bibinfo {author} {\bibfnamefont {I.}~\bibnamefont
  {Mezi\'c}},\ }\bibfield  {title} {\enquote {\bibinfo {title} {Reduction of
  three-dimensional, volume-preserving flows with symmetry},}\ }\href@noop {}
  {\bibfield  {journal} {\bibinfo  {journal} {Nonlinearity}\ }\textbf {\bibinfo
  {volume} {11}},\ \bibinfo {pages} {319--339} (\bibinfo {year}
  {1998})}\BibitemShut {NoStop}%
\bibitem [{\citenamefont {Kirby}\ and\ \citenamefont
  {Armbruster}(1992)}]{kirby_reconstructing_1992}%
  \BibitemOpen
  \bibfield  {author} {\bibinfo {author} {\bibfnamefont {M.}~\bibnamefont
  {Kirby}}\ and\ \bibinfo {author} {\bibfnamefont {D.}~\bibnamefont
  {Armbruster}},\ }\bibfield  {title} {\enquote {\bibinfo {title}
  {Reconstructing phase space from {PDE} simulations},}\ }\href@noop {}
  {\bibfield  {journal} {\bibinfo  {journal} {Z.\ Angew.\ Math.\ Phys.}\
  }\textbf {\bibinfo {volume} {43}},\ \bibinfo {pages} {999--1022} (\bibinfo
  {year} {1992})}\BibitemShut {NoStop}%
\bibitem [{\citenamefont {Bredon}(1972)}]{Bredon72}%
  \BibitemOpen
  \bibfield  {author} {\bibinfo {author} {\bibfnamefont {G.}~\bibnamefont
  {Bredon}},\ }\href@noop {} {\emph {\bibinfo {title} {Introduction to Compact
  Transformation Groups}}}\ (\bibinfo  {publisher} {Academic Press},\ \bibinfo
  {address} {New York},\ \bibinfo {year} {1972})\BibitemShut {NoStop}%
\bibitem [{\citenamefont {Palais}(1961)}]{Pal61}%
  \BibitemOpen
  \bibfield  {author} {\bibinfo {author} {\bibfnamefont {R.~S.}\ \bibnamefont
  {Palais}},\ }\bibfield  {title} {\enquote {\bibinfo {title} {On the existence
  of slices for actions of non-compact {Lie} groups},}\ }\href@noop {}
  {\bibfield  {journal} {\bibinfo  {journal} {Ann. Math.}\ }\textbf {\bibinfo
  {volume} {73}},\ \bibinfo {pages} {295--323} (\bibinfo {year}
  {1961})}\BibitemShut {NoStop}%
\bibitem [{\citenamefont {Olver}(1999)}]{OlverInv}%
  \BibitemOpen
  \bibfield  {author} {\bibinfo {author} {\bibfnamefont {P.~J.}\ \bibnamefont
  {Olver}},\ }\href@noop {} {\emph {\bibinfo {title} {Classical Invariant
  Theory}}}\ (\bibinfo  {publisher} {Cambridge Univ. Press},\ \bibinfo
  {address} {Cambridge},\ \bibinfo {year} {1999})\BibitemShut {NoStop}%
\bibitem [{\citenamefont {Rowley}\ \emph {et~al.}(2003)\citenamefont {Rowley},
  \citenamefont {Kevrekidis}, \citenamefont {Marsden},\ and\ \citenamefont
  {Lust}}]{rowley_reduction_2003}%
  \BibitemOpen
  \bibfield  {author} {\bibinfo {author} {\bibfnamefont {C.~W.}\ \bibnamefont
  {Rowley}}, \bibinfo {author} {\bibfnamefont {I.~G.}\ \bibnamefont
  {Kevrekidis}}, \bibinfo {author} {\bibfnamefont {J.~E.}\ \bibnamefont
  {Marsden}}, \ and\ \bibinfo {author} {\bibfnamefont {K.}~\bibnamefont
  {Lust}},\ }\bibfield  {title} {\enquote {\bibinfo {title} {Reduction and
  reconstruction for self-similar dynamical systems},}\ }\href@noop {}
  {\bibfield  {journal} {\bibinfo  {journal} {Nonlinearity}\ }\textbf {\bibinfo
  {volume} {16}},\ \bibinfo {pages} {1257--1275} (\bibinfo {year}
  {2003})}\BibitemShut {NoStop}%
\bibitem [{\citenamefont {Porter}\ and\ \citenamefont
  {Knobloch}(2005)}]{PoKno05}%
  \BibitemOpen
  \bibfield  {author} {\bibinfo {author} {\bibfnamefont {J.}~\bibnamefont
  {Porter}}\ and\ \bibinfo {author} {\bibfnamefont {E.}~\bibnamefont
  {Knobloch}},\ }\bibfield  {title} {\enquote {\bibinfo {title} {Dynamics in
  the 1:2 spatial resonance with broken reflection symmetry},}\ }\href
  {\doibase 10.1016/j.physd.2005.01.001} {\bibfield  {journal} {\bibinfo
  {journal} {Physica D}\ }\textbf {\bibinfo {volume} {201}},\ \bibinfo {pages}
  {318 -- 344} (\bibinfo {year} {2005})}\BibitemShut {NoStop}%
\bibitem [{\citenamefont {Dangelmayr}(1986)}]{Dang86}%
  \BibitemOpen
  \bibfield  {author} {\bibinfo {author} {\bibfnamefont {G.}~\bibnamefont
  {Dangelmayr}},\ }\bibfield  {title} {\enquote {\bibinfo {title} {Steady-state
  mode interactions in the presence of {0(2)}-symmetry},}\ }\href {\doibase
  10.1080/02681118608806011} {\bibfield  {journal} {\bibinfo  {journal} {Dyn.
  Sys.}\ }\textbf {\bibinfo {volume} {1}},\ \bibinfo {pages} {159--185}
  (\bibinfo {year} {1986})}\BibitemShut {NoStop}%
\bibitem [{\citenamefont {Armbruster}, \citenamefont {Guckenheimer},\ and\
  \citenamefont {Holmes}(1988)}]{AGHO288}%
  \BibitemOpen
  \bibfield  {author} {\bibinfo {author} {\bibfnamefont {D.}~\bibnamefont
  {Armbruster}}, \bibinfo {author} {\bibfnamefont {J.}~\bibnamefont
  {Guckenheimer}}, \ and\ \bibinfo {author} {\bibfnamefont {P.}~\bibnamefont
  {Holmes}},\ }\bibfield  {title} {\enquote {\bibinfo {title} {Heteroclinic
  cycles and modulated travelling waves in systems with {O(2)} symmetry},}\
  }\href@noop {} {\bibfield  {journal} {\bibinfo  {journal} {Physica D}\
  }\textbf {\bibinfo {volume} {29}},\ \bibinfo {pages} {257--282} (\bibinfo
  {year} {1988})}\BibitemShut {NoStop}%
\bibitem [{\citenamefont {Jones}\ and\ \citenamefont
  {Proctor}(1987)}]{JoPro87}%
  \BibitemOpen
  \bibfield  {author} {\bibinfo {author} {\bibfnamefont {C.~A.}\ \bibnamefont
  {Jones}}\ and\ \bibinfo {author} {\bibfnamefont {M.~R.~E.}\ \bibnamefont
  {Proctor}},\ }\bibfield  {title} {\enquote {\bibinfo {title} {Strong spatial
  resonance and travelling waves in {Benard} convection},}\ }\href {\doibase
  10.1016/0375-9601(87)90008-9} {\bibfield  {journal} {\bibinfo  {journal}
  {Phys. Lett. A}\ }\textbf {\bibinfo {volume} {121}},\ \bibinfo {pages}
  {224--228} (\bibinfo {year} {1987})}\BibitemShut {NoStop}%
\bibitem [{\citenamefont {Golubitsky}, \citenamefont {Stewart},\ and\
  \citenamefont {Schaeffer}(1988)}]{golubII}%
  \BibitemOpen
  \bibfield  {author} {\bibinfo {author} {\bibfnamefont {M.}~\bibnamefont
  {Golubitsky}}, \bibinfo {author} {\bibfnamefont {I.}~\bibnamefont {Stewart}},
  \ and\ \bibinfo {author} {\bibfnamefont {D.~G.}\ \bibnamefont {Schaeffer}},\
  }\href@noop {} {\emph {\bibinfo {title} {Singularities and Groups in
  Bifurcation Theory, vol. II}}}\ (\bibinfo  {publisher} {Springer},\ \bibinfo
  {address} {New York},\ \bibinfo {year} {1988})\BibitemShut {NoStop}%
\bibitem [{\citenamefont {Lorenz}(1963)}]{lorenz}%
  \BibitemOpen
  \bibfield  {author} {\bibinfo {author} {\bibfnamefont {E.~N.}\ \bibnamefont
  {Lorenz}},\ }\bibfield  {title} {\enquote {\bibinfo {title} {Deterministic
  nonperiodic flow},}\ }\href@noop {} {\bibfield  {journal} {\bibinfo
  {journal} {J. Atmos. Sci.}\ }\textbf {\bibinfo {volume} {20}},\ \bibinfo
  {pages} {130--141} (\bibinfo {year} {1963})}\BibitemShut {NoStop}%
\bibitem [{\citenamefont {H\'enon}(1976)}]{henon}%
  \BibitemOpen
  \bibfield  {author} {\bibinfo {author} {\bibfnamefont {M.}~\bibnamefont
  {H\'enon}},\ }\bibfield  {title} {\enquote {\bibinfo {title} {A
  two-dimensional mapping with a strange attractor},}\ }\href@noop {}
  {\bibfield  {journal} {\bibinfo  {journal} {Commun. Math. Phys.}\ }\textbf
  {\bibinfo {volume} {50}},\ \bibinfo {pages} {69} (\bibinfo {year}
  {1976})}\BibitemShut {NoStop}%
\bibitem [{\citenamefont {R\"ossler}(1976)}]{ross}%
  \BibitemOpen
  \bibfield  {author} {\bibinfo {author} {\bibfnamefont {O.~E.}\ \bibnamefont
  {R\"ossler}},\ }\bibfield  {title} {\enquote {\bibinfo {title} {An equation
  for continuous chaos},}\ }\href@noop {} {\bibfield  {journal} {\bibinfo
  {journal} {Phys. Lett. A}\ }\textbf {\bibinfo {volume} {57}},\ \bibinfo
  {pages} {397} (\bibinfo {year} {1976})}\BibitemShut {NoStop}%
\bibitem [{\citenamefont {Devaney}(1989)}]{devnmap}%
  \BibitemOpen
  \bibfield  {author} {\bibinfo {author} {\bibfnamefont {R.~L.}\ \bibnamefont
  {Devaney}},\ }\href@noop {} {\emph {\bibinfo {title} {An Introduction to
  Chaotic Dynamical systems}}}\ (\bibinfo  {publisher} {Wesley},\ \bibinfo
  {address} {Redwood City},\ \bibinfo {year} {1989})\BibitemShut {NoStop}%
\bibitem [{\citenamefont {Cvitanovi\'{c}}(2007)}]{Cvi07}%
  \BibitemOpen
  \bibfield  {author} {\bibinfo {author} {\bibfnamefont {P.}~\bibnamefont
  {Cvitanovi\'{c}}},\ }\href@noop {} {\enquote {\bibinfo {title} {Continuous
  symmetry reduced trace formulas},}\ } (\bibinfo {year} {2007}),\ \bibinfo
  {note} {\\ \wwwcb{/$\sim$predrag/papers/trace.pdf}}\BibitemShut {NoStop}%
\bibitem [{\citenamefont {Cvitanovi\'{c}}\ and\ \citenamefont
  {Eckhardt}(1991)}]{pexp}%
  \BibitemOpen
  \bibfield  {author} {\bibinfo {author} {\bibfnamefont {P.}~\bibnamefont
  {Cvitanovi\'{c}}}\ and\ \bibinfo {author} {\bibfnamefont {B.}~\bibnamefont
  {Eckhardt}},\ }\bibfield  {title} {\enquote {\bibinfo {title} {Periodic orbit
  expansions for classical smooth flows},}\ }\href@noop {} {\bibfield
  {journal} {\bibinfo  {journal} {J. Phys. A}\ }\textbf {\bibinfo {volume}
  {24}},\ \bibinfo {pages} {L237} (\bibinfo {year} {1991})}\BibitemShut
  {NoStop}%
\bibitem [{\citenamefont {Cvitanovi\'{c}}\ and\ \citenamefont
  {Eckhardt}(1993)}]{CvitaEckardt}%
  \BibitemOpen
  \bibfield  {author} {\bibinfo {author} {\bibfnamefont {P.}~\bibnamefont
  {Cvitanovi\'{c}}}\ and\ \bibinfo {author} {\bibfnamefont {B.}~\bibnamefont
  {Eckhardt}},\ }\bibfield  {title} {\enquote {\bibinfo {title} {Symmetry
  decomposition of chaotic dynamics},}\ }\href@noop {} {\bibfield  {journal}
  {\bibinfo  {journal} {Nonlinearity}\ }\textbf {\bibinfo {volume} {6}},\
  \bibinfo {pages} {277--311} (\bibinfo {year} {1993})},\ \bibinfo {note}
  {\arXiv{chao-dyn/9303016}}\BibitemShut {NoStop}%
\bibitem [{\citenamefont {Creagh}(1993)}]{Creagh93}%
  \BibitemOpen
  \bibfield  {author} {\bibinfo {author} {\bibfnamefont {S.~C.}\ \bibnamefont
  {Creagh}},\ }\bibfield  {title} {\enquote {\bibinfo {title} {Semiclassical
  mechanics of symmetry reduction},}\ }\href@noop {} {\bibfield  {journal}
  {\bibinfo  {journal} {J. Phys. A}\ }\textbf {\bibinfo {volume} {26}},\
  \bibinfo {pages} {95--118} (\bibinfo {year} {1993})}\BibitemShut {NoStop}%
\bibitem [{\citenamefont {Artuso}, \citenamefont {Aurell},\ and\ \citenamefont
  {Cvitanovi{\'{c}}}(1990)}]{AACI}%
  \BibitemOpen
  \bibfield  {author} {\bibinfo {author} {\bibfnamefont {R.}~\bibnamefont
  {Artuso}}, \bibinfo {author} {\bibfnamefont {E.}~\bibnamefont {Aurell}}, \
  and\ \bibinfo {author} {\bibfnamefont {P.}~\bibnamefont {Cvitanovi{\'{c}}}},\
  }\bibfield  {title} {\enquote {\bibinfo {title} {Recycling of strange sets:
  {I}. {Cycle} expansions},}\ }\href@noop {} {\bibfield  {journal} {\bibinfo
  {journal} {Nonlinearity}\ }\textbf {\bibinfo {volume} {3}},\ \bibinfo {pages}
  {325--359} (\bibinfo {year} {1990})}\BibitemShut {NoStop}%
\bibitem [{\citenamefont {Rugh}(1992)}]{hhrugh92}%
  \BibitemOpen
  \bibfield  {author} {\bibinfo {author} {\bibfnamefont {H.~H.}\ \bibnamefont
  {Rugh}},\ }\bibfield  {title} {\enquote {\bibinfo {title} {The correlation
  spectrum for hyperbolic analytic maps},}\ }\href {\doibase
  10.1088/0951-7715/5/6/003} {\bibfield  {journal} {\bibinfo  {journal}
  {Nonlinearity}\ }\textbf {\bibinfo {volume} {5}},\ \bibinfo {pages} {1237}
  (\bibinfo {year} {1992})}\BibitemShut {NoStop}%
\bibitem [{\citenamefont {Wolf}\ \emph {et~al.}(1985)\citenamefont {Wolf},
  \citenamefont {Swift}, \citenamefont {Swinney},\ and\ \citenamefont
  {Vastano}}]{WolfSwift85}%
  \BibitemOpen
  \bibfield  {author} {\bibinfo {author} {\bibfnamefont {A.}~\bibnamefont
  {Wolf}}, \bibinfo {author} {\bibfnamefont {J.~B.}\ \bibnamefont {Swift}},
  \bibinfo {author} {\bibfnamefont {H.~L.}\ \bibnamefont {Swinney}}, \ and\
  \bibinfo {author} {\bibfnamefont {J.~A.}\ \bibnamefont {Vastano}},\
  }\bibfield  {title} {\enquote {\bibinfo {title} {Determining {Lyapunov}
  exponents from a time series},}\ }\href@noop {} {\bibfield  {journal}
  {\bibinfo  {journal} {Physica D}\ }\textbf {\bibinfo {volume} {16}},\
  \bibinfo {pages} {285--317} (\bibinfo {year} {1985})}\BibitemShut {NoStop}%
\bibitem [{\citenamefont {Cvitanovi{\'c}}, \citenamefont {Davidchack},\ and\
  \citenamefont {Siminos}(2010)}]{SCD07}%
  \BibitemOpen
  \bibfield  {author} {\bibinfo {author} {\bibfnamefont {P.}~\bibnamefont
  {Cvitanovi{\'c}}}, \bibinfo {author} {\bibfnamefont {R.~L.}\ \bibnamefont
  {Davidchack}}, \ and\ \bibinfo {author} {\bibfnamefont {E.}~\bibnamefont
  {Siminos}},\ }\bibfield  {title} {\enquote {\bibinfo {title} {On the state
  space geometry of the {Kuramoto-Sivashinsky} flow in a periodic domain},}\
  }\href@noop {} {\bibfield  {journal} {\bibinfo  {journal} {SIAM J. Appl. Dyn.
  Syst.}\ }\textbf {\bibinfo {volume} {9}},\ \bibinfo {pages} {1--33} (\bibinfo
  {year} {2010})},\ \bibinfo {note} {\arXiv{0709.2944}}\BibitemShut {NoStop}%
\bibitem [{\citenamefont {Lan}\ and\ \citenamefont
  {Cvitanovi{\'c}}(2008)}]{lanCvit07}%
  \BibitemOpen
  \bibfield  {author} {\bibinfo {author} {\bibfnamefont {Y.}~\bibnamefont
  {Lan}}\ and\ \bibinfo {author} {\bibfnamefont {P.}~\bibnamefont
  {Cvitanovi{\'c}}},\ }\bibfield  {title} {\enquote {\bibinfo {title} {Unstable
  recurrent patterns in {Kuramoto-Sivashinsky} dynamics},}\ }\href@noop {}
  {\bibfield  {journal} {\bibinfo  {journal} {Phys. Rev. E}\ }\textbf {\bibinfo
  {volume} {78}},\ \bibinfo {pages} {026208} (\bibinfo {year} {2008})},\
  \bibinfo {note} {\arXiv{0804.2474}}\BibitemShut {NoStop}%
\bibitem [{\citenamefont {Dettmann}\ and\ \citenamefont
  {Morriss}(1997)}]{DM97}%
  \BibitemOpen
  \bibfield  {author} {\bibinfo {author} {\bibfnamefont {C.~P.}\ \bibnamefont
  {Dettmann}}\ and\ \bibinfo {author} {\bibfnamefont {G.~P.}\ \bibnamefont
  {Morriss}},\ }\bibfield  {title} {\enquote {\bibinfo {title} {Stability
  ordering of cycle expansions},}\ }\href {\doibase
  10.1103/PhysRevLett.78.4201} {\bibfield  {journal} {\bibinfo  {journal}
  {Phys. Rev. Lett.}\ }\textbf {\bibinfo {volume} {78}},\ \bibinfo {pages}
  {4201--4204} (\bibinfo {year} {1997})}\BibitemShut {NoStop}%
\bibitem [{\citenamefont {Ding}\ and\ \citenamefont
  {Cvitanovi\'c}(2014)}]{DingCvit14}%
  \BibitemOpen
  \bibfield  {author} {\bibinfo {author} {\bibfnamefont {X.}~\bibnamefont
  {Ding}}\ and\ \bibinfo {author} {\bibfnamefont {P.}~\bibnamefont
  {Cvitanovi\'c}},\ }\href@noop {} {\enquote {\bibinfo {title} {Periodic
  eigendecomposition and its application in {Kuramoto-Sivashinsky} system},}\ }
  (\bibinfo {year} {2014}),\ \bibinfo {note} {\arXiv{1406.4885}}\BibitemShut
  {NoStop}%
\end{thebibliography}%

\end{document}